\documentclass[lettersize,journal]{IEEEtran}
\usepackage{amsmath,amssymb,amsfonts}
\usepackage{array}
\usepackage[caption=false,font=normalsize,labelfont=sf,textfont=sf]{subfig}
\usepackage{textcomp}
\usepackage{stfloats}
\usepackage[skip=10pt]{caption}
\usepackage{url}
\usepackage{verbatim}
\usepackage{makecell}
\usepackage{booktabs}
\usepackage[table,xcdraw]{xcolor}
\usepackage{textcomp}
\usepackage{multirow}
\usepackage{array}
\usepackage{graphicx}
\usepackage{cite}
\usepackage[export]{adjustbox}
\usepackage{algorithm}
\usepackage{algpseudocode}
\usepackage{capt-of}

\begin{document}

\title{TRANSPOSE: Transitional Approaches for Spatially-Aware LFI Resilient FSM Encoding}

\author{\IEEEauthorblockN{Muhtadi Choudhury\IEEEauthorrefmark{1}, Minyan Gao\IEEEauthorrefmark{1}, Avinash Varna\IEEEauthorrefmark{4}, Elad Peer\IEEEauthorrefmark{5} and Domenic Forte\IEEEauthorrefmark{1} \\}
\IEEEauthorblockA{\IEEEauthorrefmark{1} University of Florida, Gainesville, FL, USA\\}
\IEEEauthorblockA{\IEEEauthorrefmark{4} Intel Corporation, USA}
\IEEEauthorblockA{\IEEEauthorrefmark{5} Intel Corporation, Israel}
}

\maketitle

\begin{abstract}

Finite state machines (FSMs) regulate sequential circuits, including access to sensitive information and privileged CPU states. Courtesy of contemporary research on laser attacks, laser-based fault injection (LFI) is becoming even more precise
where an adversary can thwart chip security by altering individual flip-flop (FF) values. Different laser models, e.g., bit flip, bit set, and bit reset, have been developed to appreciate LFI on practical targets. As traditional approaches may incorporate substantial overhead, state-based SPARSE and transition-based TAMED countermeasures were proposed in our prior work to improve FSM resiliency efficiently. TAMED overcame SPARSE’s limitation of being too conservative, and generating multiple LFI resilient encodings for contemporary LFI models on demand. SPARSE, however, incorporated design layout information into its vulnerability estimation which makes its vulnerability estimation metric more accurate. In this paper, we extend TAMED by proposing a transition-based encoding CAD framework (TRANSPOSE), that incorporates spatial transitional vulnerability metrics to quantify design susceptibility of FSMs based on both the bit flip model and the set-reset models. TRANSPOSE also incorporates floorplan optimization into its framework to accommodate secure spatial inter-distance of FF-sensitive
regions. All TRANSPOSE approaches are demonstrated on 5 multifarious benchmarks and outperform existing FSM encoding schemes/frameworks in terms of security and overhead.

\end{abstract}


\section{Introduction}

Physical attacks can target secure portions of system on chips (SoCs) and cryptographic circuits thereby jeopardizing the integrity and confidentiality. Among the options, fault injection attacks entail external or internal \emph{active} maneuvering that lead to a fault. Laser fault injection (LFI) stands out as a highly precise method capable of inducing faults at a very fine resolution (even affecting just a single byte or bit)~\cite{agoyan2010flip}. A contemporary LFI set-up allows control of fault injection time (pulse duration and shot instant), repeatability, and localization.
Unlike other fault attacks like voltage variations~\cite{blomer2003fault} or clock glitches~\cite{agoyan2010clocks}, LFI requires strict adherence to specific constraints regarding duration to ensure both spatial and temporal accuracy, thereby ensuring an exact fault occurs~\cite{ schellenberg2016large, leveugle2014laser}. 
Experiments on LFI demonstrate \emph{data dependent and data independent fault models}, i.e., bit-reset/set models and bit flip model, respectively~\cite{roscian2013frontside}.
A bit reset (resp. a bit-set) models a fault that alters the target bit from 1 to 0 (resp. from 0 to 1). However, if the current bit is already at 0 (resp. 1 for bit-set) there is no effect. A bit-flip corresponds to a fault irrespective of the target's current state.

Current research highlights the \emph{laser-sensitive areas} in a D flip-flop (DFF) to laser-induced faults, considering both data-dependent and data-independent fault models~\cite{champeix2015seu, dutertre2018case}. Attackers can exploit these precise vulnerable regions in current and future technology nodes~\cite{dutertre2018case}.  
It is crucial to acknowledge the significance of identifying these specific sensitive areas when developing countermeasures against LFI. Even targeting a few transistors with a less precise/ relaxed laser spot can cause significant faults, highlighting the absence of inherent protection against LFI even at the nanoscale technology level~\cite{schellenberg2016large}.
Furthermore, as effectuating faults with relaxed spot size is a possibility, countermeasures must also incorporate as many relaxed constraints on DFFs as possible to conserve area and power along with security still intact~\cite{choudhury2021sparse}.

Current countermeasures such as hardware irradiation detectors~\cite{leveugle2014laser} are costly. CAD tools, in contrast, can automatically integrate logical methods such as redundancy or security-aware encoding techniques~\cite{nahiyan2016avfsm}. Another strategy involves \emph{state exploration} using coding theory approaches, where each state of a finite state machine (FSM) is treated as a linear or nonlinear code. This enhances FSM resilience against fault injection (FI) through error correction or detection mechanisms. However, this approach increases chip area, power usage, and affects performance since it assumes \emph{all states require equal protection}. 
In our earlier research, we introduced state exploration methods that promote LFI-resistant encoding for arbitrary FSM sizes and numbers of lasers, specifically PATRON and SPARSE~\cite{choudhury2021patron, choudhury2021sparse}. Unlike approaches based solely on coding theory, these methods focus on safeguarding critical FSM states, making them less conservative. However, they do not take into account the exact laser-sensitive areas necessary to distinguish between data-dependent bit-set/reset and data-independent bit-flip fault models. Additionally, PATRON and SPARSE assume protection for all sensitive states in the FSM, which can prove to be weighty in the overhead. To that end, a more recent transition-based approach named TAMED is proposed~\cite{itcpaper} which protects only the specific \emph{authorized transitions}~\cite{nahiyan2016avfsm} in the FSM.

Although all the above approaches individually provide unique elements that benefit LFI research, an all-encompassing comprehensive framework is missing in the literature that cherry-picks all the beneficial concepts and combines them into an efficient vulnerability-monitoring and low overhead approach. Further, advanced architectures are always required to optimize cost, area, performance penalties, and power consumption as they are crucial constraints in modern system design. Thus, examining all the contrasting design requirements while maintaining proper security, we introduce TRANSPOSE (\underline{TRANS}itional A\underline{P}proaches f\underline{O}r \underline{S}patially-Aware LFI Resilient State Machine \underline{E}ncoding) which makes the FSM inherently tolerant to precise LFI sensitive areas. 
Particularly, our contributions are:


\begin{itemize}
\item An \emph{automated} generation of LFI-resistant state encoding that integrates with commercial CAD tools such as Design Compiler and IC Compiler II. Through the use of linear programming (LP), TRANSPOSE can identify a single, LFI-resistant encoding \emph{without any manual input}.

\item We propose the \emph{Spatial Transitional Vulnerability Metrics} ($STVM$), 
which identify vulnerabilities missed by the previously proposed $VM$ (PATRON), $SVM$ (SPARSE), and $TVM$ (TAMED). $STVM$ incorporate 
FF-sensitive regions and thus address both data-dependent and data-independent models.

\item We expand TRANSPOSE's LP criteria to \emph{protect as many critical transitions from both the data-dependent and data-independent models}. For any arbitrary FSM, encoding, and placement are co-optimized in terms of area overhead, switching activity (dynamic power consumption), and security for a multi-laser adversary.



\item We demonstrate TRANSPOSE on 5 diverse controller benchmarks and compare its security and overhead 
to other security-aware encoding techniques. 

\end{itemize}

An outline for the rest of this paper is as follows.  
In the next section, we discuss basic notation and common terms, FSM definitions, and a motivating example. 
In Section~\ref{LFITM}, security assessment via contemporary work and flip-flop sensitivity are described and used to constitute a realistic threat model for TRANSPOSE. Subsequently, examples with previously proposed metrics are shown to misconstrue vulnerability in FSMs triggering the need for $STVM$.  
Section~\ref{TMe} delineates the TRANSPOSE methodology which incorporates a precise model, discussion on salient parameters, and multiple transition types along with the proposed metric leading to comprehensive secure encoding and floorplan optimization procedures. Results are presented and discussed in Section~\ref{randd}. Finally, conclusions and future work are given in the last section.

\vspace*{-0.5cm}
\section{Background Concepts}

\begin{table*}[t]
 \centering
  \scriptsize
  \caption{\textcolor{blue}{Terms and definitions for variables, sets, metrics, and ILP formulation.}} 
 \tabcolsep=0.08cm
\begin{tabular}{|c|c|c|c|c|c|}
\toprule
\textbf{Term}  & \textbf{Definition}                            & \textbf{Term}              & \textbf{Definition}                                                                                  & \textbf{Term} & \textbf{Definition}                              \\
\midrule  

\multicolumn{6}{|c|}{\textbf{General Variables and Terms}}                                                                                                                                                                \\

\midrule  

$x$     & \# of lasers                         & $f$            & Number of injected fault events                                       & $D$                 & Diameter of laser   beam \\
$n$     & Total \#  of flip-flops in FSM          & $\vec{l}_i$  &  Coordinate vector of the $i$th laser                                                                   & $HD$   & Hamming   distance                     \\
$p$   & Transitional probability                      & $L$               & Laser location matrix                                                                        &  $FF$    &  Flip-flop                                       \\

\midrule 
\multicolumn{6}{|c|}{\textbf{Sets and Metrics}}                                                           \\

\midrule 
$\mathbb{E}(x)$ & Vulnerable FFs  assuming $x$ lasers & $\mathbb{SVT}(x)$            & Spatially vulnerable transitions assuming $x$ lasers                                               & $\mathbb{S}$    & FSM   states                            \\
$\mathbb{P}$     & Protected   states                    & $\mathbb{AU}$                & Authorized states                                                                           & $\mathbb{FF}$   & Flip-flops                            \\
$\mathbb{NS}$    & Normal states                         & $\mathbb{SS}$                & Sensitive states                                                                          &  $\mathbb{NFF}$  & Normal flip-flops                       \\
$\mathbb{V}$     & Vulnerable FF combinations w.r.t. fault types       & $\mathcal{P}(\mathbb{FF})$    & All possible collections of FFs                                                         &  $\mathbb{SFF}$  & Secure flip-flops                    \\
$VM$    & Vulnerability    metric               & $SVM$               & Spatial vulnerability metric                                                              &  $\mathbb{SVT}$    &   Spatially vulnerable
transitions                \\
$TVM$    & Transitional vulnerability    metric               & $STVM$               & Spatial transitional vulnerability metric                                                              &  $\mathbb{AT}$    &   Authorized transitions                \\

\midrule 
\multicolumn{6}{|c|}{\textbf{ILP Terms}}                                                                                                                                                                                    \\
\midrule 
$W$     & Upper bound of floorplan width        & $H$                 & Upper   bound of floorplan height                                                           & $h_i$  & Height of $FF_i$                       \\
$Y$     & Optimal floorplan height              & $y_{ij}$,    $z_{ij}$ & Binary variables for relative position of $FF_i$ vs. $FF_j$ & $w_i$  & Width   of $FF_i$     \\
\bottomrule
\end{tabular}
\label{tab:terms}
\end{table*}

\subsection{Basic Notation and Common Terms}
Blackboard bold font with upper case letters, e.g., $\mathbb{S}$, is used to represent sets. Upper case and italic font with one or no subscripts signify an element of a set, e.g., $S_i$ or $S$. 
Vectors are denoted using lower-case with arrow, e.g., $\vec{v}$, while vector elements are written in lower-case with a subscript, e.g., $v_i$. The shorthand notation for a subset of elements $i$ to $j$ of a vector $\vec{v}$ is  $\vec{v}_{i:j}$. A relation between the $i$th and $j$th elements is denoted with a subscript ${ij}$. The operator $|\cdot|$ denotes the cardinality of a set. 
For the reader's reference, we provide Table~\ref{tab:terms} which contains the terms and brief descriptions.
\vspace*{-0.3cm}

\subsection{FSM and Encoding} \label{FSMencoding}

An FSM is defined as a 5-tuple  
$
  ( \mathbb{S}, \mathbb{I}, \mathbb{O}, \varphi,\lambda)
$,
where $\mathbb{S}$ is a finite set of states, $\mathbb{I}$ is a finite set of input symbols, $\mathbb{O}$ is a finite set of output symbols, $\varphi$ is the next-state function and $\lambda$ is the output function. 
Typically, an FSM is depicted as a directed graph $\mathcal{G} = (\mathbb{S}, \mathbb{T}) $ where each state $ S \in \mathbb{S} $ represents a vertex and each edge $T_{ij} \in \mathbb{T}$ represents a transition or edge from state $S_i$ to the state $S_j$. 

Each state in an FSM is only to be admitted from its \emph{accessible set of states}, i.e., 
$
A(S_j) = \{ S_i\: |\: T_{ij} \in \mathbb{T}\}
$. 
In~\cite{nahiyan2016avfsm}, a designer designates a set $\mathbb{P}$ of \emph{protected states} and a set $\mathbb{AU}$ of \emph{authorized states}. A transition from state $S \in \mathbb{AU}$ that is allowed access to $\mathbb{P}$, such that \textcolor{blue}{$ A(P) ={\{ P\: |\: P \in \mathbb{P} \}}$ }is referred to as an \emph{authorized transition} ($\mathbb{AT}$) in this paper. To put it differently, when the current state is an authorized state and the next state is the protected state, authorized transitions manifest; the direction of the edge in $AT$ is always from $\mathbb{AU}$ to $\mathbb{P}$. 
In recent work relating to LFI resilient FSMs that consider only bit flip model~\cite{choudhury2021patron, choudhury2021sparse}, a state exploration scheme is chosen where \emph{all} \emph{normal states} ($\mathbb{NS}$) are secure from the \emph{sensitive states} ($\mathbb{SS}$) defined as $\mathbb{NS} = \{ S \in \mathbb{S} \: |\:  s \notin \mathbb{AU} \cup \mathbb{P} \}$, $\mathbb{SS} = \{ S \in \mathbb{S} \: |\:  s \in \mathbb{AU} \cup \mathbb{P} \}$, respectively. In other words, $\mathbb{NS}$ is extrinsic as far as $\mathbb{AT}$ is concerned.


\subsection{Fault Injection Against FSMs}

\begin{figure}[t]
  \centering
  \includegraphics[width=0.60\linewidth]{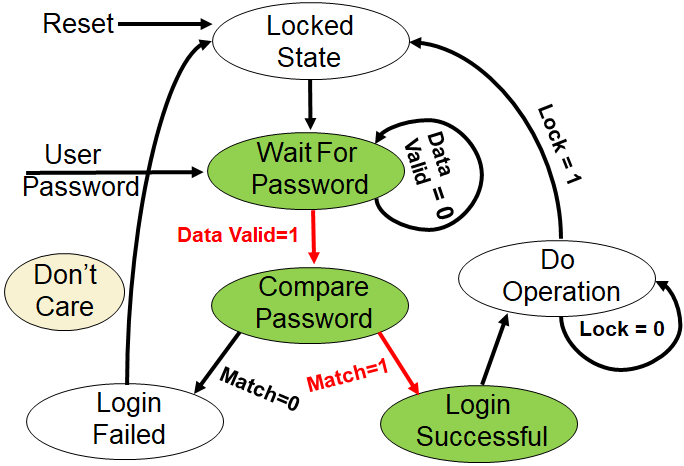}
  \caption{A password checking FSM where the fault causes an incorrect password to be accepted.}
  \label{fsmintro}
\end{figure}

The effectiveness of fault injection attacks is depicted using a basic state transition diagram of a password authentication FSM in Fig.~\ref{fsmintro}. The FSM constitutes 6 states: `Locked State', `Compare Password', `Wait For Password', `Login Failed', `Login Successful', and `Do Operation'. After reset, the system launches in Locked State. In the next clock-rising edge, the system advances to Wait For Password and is held there until the user inputs a password. Compare Password evaluates the user password with system password. The premise is that only authorized users possess the correct password. If passwords mismatch, the FSM moves to Login Failed and later reverts to Locked State. When passwords match, the user successfully accesses the system and advances to Do Operation. The user can choose to return the FSM to Locked State at a later point.

The primary objective of this FSM is to prevent any malicious users from circumventing the password comparison process and advancing to Login Successful, which grants system access. Therefore, the most crucial transition for the FSM designer is the one associated with `Match=1'. Additionally, the designer must ensure that the FSM moves to Compare Password for each user to verify their inputs; skipping this step could potentially facilitate unauthorized access to the system. Consequently, any FI that successfully triggers these two transitions would enable an attacker to bypass the authentication mechanism established by the protocol. 

\subsection{D Flip-Flop Operation} 
\begin{figure}[t]
  \centering
  \includegraphics[width=0.75\linewidth]{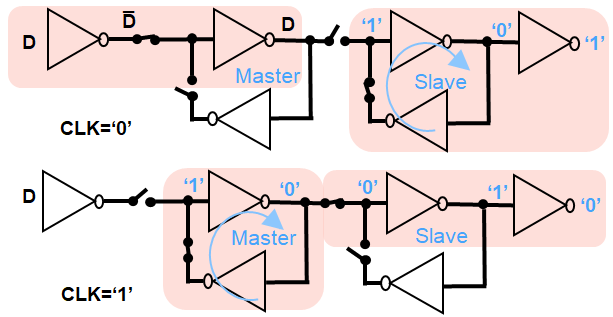}
  \caption{D flip-flop operation for logic low and high clock signal. Orange regions highlight the active circuits and blue semi-circle indicates a latch in hold mode.}
  \label{fig:DFF}
\end{figure}

A common method to construct an edge-triggered D flip-flop (FF) is employing a master-slave latch configuration~\cite{rabaey2003digital} as shown in Fig.~\ref{fig:DFF}. When the clock signal is low ($CLK=0$), the input $D$ directly reflects its output. Simultaneously, the slave latch maintains its previous value at the FF output ($Q$) through positive feedback in hold mode. Upon the clock transitioning to logic high ($CLK=1$), the master and slave latches switch roles to hold and transparent modes, respectively. Consequently, the FF output $Q$ adopts the input $D$'s most recent value prior to the clock's rising edge.

\section{Threat model, Flip-Flop Sensitivity and Security Assessment} \label{LFITM}

\subsection{Proposed Threat Model}

Our threat model's scope is defined using comprehensive definitions and updated fault models from recent research sources~\cite{richter2021fiver, richter2021revisiting,choudhury2023enhanced}. Circuits are categorized into combinational logic gates ($\mathbb{g}_{cm}$) and state elements ($\mathbb{g}_s=\{FF\}$). An attacker is characterized by a function $\zeta(f,t,l)$, where $f$ represents the total number of fault events (spatial and temporal components), $t$ describes fault types (bit flip, reset, set), and $l$ denotes fault location(s) in digital logic circuits.
A flip-flop ($FF$) in the circuit experiences a bit set (or reset) if it transitions exclusively from state 0 to state 1 (or from state 1 to state 0), while the bit flip model involves inverting the $FF$ value. When considering $f$, spatial or temporal dimensions (\emph{univariate} or \emph{multivariate}) must be taken into account. Univariate fault injections involve fault events within the same clock cycle, whereas multivariate fault injections occur across different clock cycles.

Laser-induced faults in state elements occur through \emph{Single Event Transient (SET)} and \emph{Single Event Upset (SEU)}. In SET, the laser targets the combinational logic section, and a fault transient propagates to the memory cell within the memorization time window. In SEU, the laser directly strikes the memory cell, causing an immediate bit-flip without delay.
The component influencing transient fault success rates, specifically fault propagation through combinational logic, is termed as \emph{masking}~\cite{miskov2010modeling, shivakumar2002modeling}. There are three types of masking: electrical, logical, and latching-window, each of which inhibits fault propagation to flip-flop (FF) inputs by attenuating faults, controlling inputs, and adjusting memorization time windows, respectively. For further details, readers are directed to~\cite{miskov2008process, peng2009soft}.

From the aforementioned discussion, Single Event Upset (SEU) is resilient against these masking mechanisms. Therefore, this paper focuses on direct memory units as targets, specifically state elements represented as $\mathbb{g}=\mathbb{g}_s=\{FF\}$.
Our assumptions regarding the target, FSM knowledge, fault types, number of concurrent faults, and their locations align with those outlined in~\cite{itcpaper}. In summary, this paper assumes the attack model $\zeta(f,\tau_{set-reset/bf},\mathbb{g})$~\cite{richter2021revisiting}, where $f$ denotes the number of univariate faults induced by $x$ laser beams within a single clock cycle, considering set-reset or bit-flip models at any state FF locations in the design.

\begin{figure}[t]
  \centering
  \includegraphics[width=0.80\linewidth]{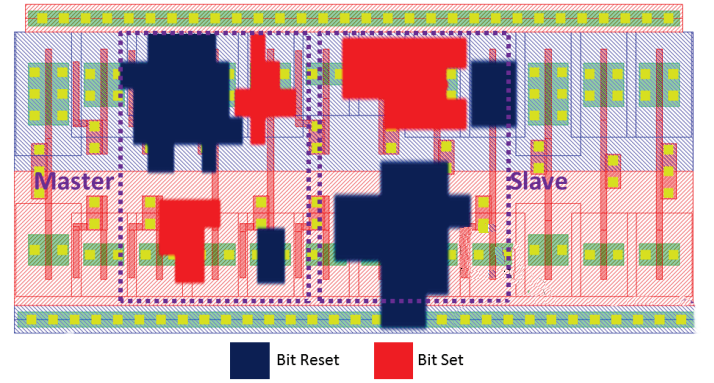}
  \caption{Experimental results showing the sensitivity map on a D Flip-Flop with laser stimulation~\cite{champeix2015seu}.}
  \label{proto}
\end{figure}

\subsection{SEU Sensitive Regions of Flip-Flop} 

The occurrence of Single Event Upset (SEU) hinges on numerous attack parameters, such as laser spot size, FF's sensitive areas, power levels, pulse duration, spatial characteristics (location, geometry, wafer thickness), and PN junction voltage biasing.

The latest electrical model, addressing ultra-short laser pulses and precise spatial resolution to pinpoint sensitive areas in current CMOS technology, is detailed in~\cite{champeix2015seu}. These areas, highly susceptible to laser impacts, are identified through cartographic measurements and validated by rigorous electrical simulations that consider the target's topology, as depicted in Fig.~\ref{proto}. We discuss how the sensitivity map influences critical decisions in assessing FSM vulnerabilities, drawing on various methodologies from the literature in Section~\ref{saotbabfm}.

\subsection{Transitional vs State-based Protection Approaches}

In this section, we examine how the newly introduced transitional methodologies (TAMED)~\cite{itcpaper}, which include data-independent and data-dependent principles, address FSM security in contrast to state-based protection methods. Transitional methodologies specifically target $\mathbb{AT}$ by addressing various sensitive areas of $FF$, such as bit set and bit reset models. In contrast, state-based approaches focus on countermeasures that involve state exploration strategies related to $\mathbb{SS}$ and the bit flip model.

\subsubsection{Defining Precise Set-Reset Model}

\setlength\intextsep{2pt}
\begin{figure}[t]
  \centering
  \includegraphics[width=0.85\linewidth]{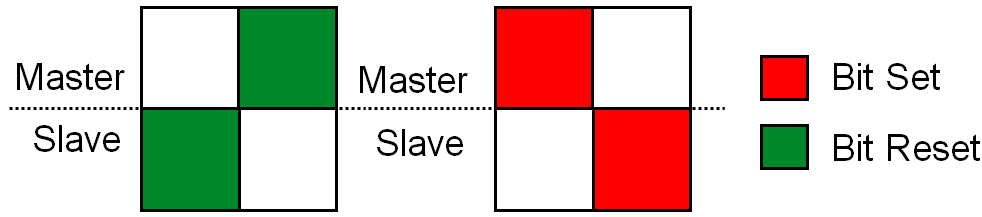}
  \caption{Simplified representation of sensitive areas in a DFF.
  Sensitive areas for a bit reset (left) and for a bit set (right).}
  \label{tolopology}
\end{figure}

The topological sensitivity outcomes from laser stimulation experiments on a DFF~\cite{champeix2015seu} are depicted in Fig.~\ref{tolopology} for clarity. The sensitivity map illustrates approximate vulnerable areas within the master and slave latches for both bit reset and bit set. This indicates that only a precise laser beam targeted at these sensitive regions can induce an instantaneous change from `$1$' to `$0$' (or `$0$' to `$1$') resulting in a Single Event Upset (SEU). This approach is referred to as the more precise set-reset model. However, it also suggests that the assumption of instantaneous state reversal from `$1$' to `$0$' (or vice versa) across any part of the $FF$ layout in the bit flip model may not always hold, as we discuss below. Note that, any appropriate countermeasure proposed must provide flexibility to adjust these sensitive FF regions as the sensitive regions may vary for different types of FFs~\cite{kauppila2011impact}.

\subsubsection{Security Assessment of Transition- vs. State-based Approaches} \label{saotbabfm}

\begin{table}[]
\centering
\tiny
\begin{tabular}{|c|c|l}
\cline{1-2}
\cellcolor[HTML]{FFCCC9}\textbf{Previously Proposed   Metrics}                                                                                                                   & \cellcolor[HTML]{FFCCC9}\textbf{Limitations}                                                                                        &  \\ \cline{1-2}
\begin{tabular}[c]{@{}c@{}}$VM$ is the percentage \\ of states where $x$ laser faults can access a sensitive  state\end{tabular}                                   & \begin{tabular}[c]{@{}c@{}}\textcolor{blue}{Only bit flip model;  transition order, FF sensitive} \\\textcolor{blue}{regions and secure FF placements unaddressed}\end{tabular} &  \\ \cline{1-2}
\begin{tabular}[c]{@{}c@{}}$SVM$   is the percentage of states where\\ $x$ faults can lead to a $SS$ considering   $\mathbb{FF}$ layout\end{tabular}         & \begin{tabular}[c]{@{}c@{}}\textcolor{blue}{Only bit flip model;   transition order and}\\ \textcolor{blue}{FF sensitive regions unaddressed}\end{tabular}                          &  \\ \cline{1-2}
\begin{tabular}[c]{@{}c@{}}$TVM$  is the percentage of states where \\ $x$ bit flip or set-reset faults can lead to an authorized transition\end{tabular} & \textcolor{blue}{Secure FF placements unaddressed}                                                                                      &  \\ \cline{1-2}
\end{tabular}%
\caption{Limitations of pertinent security metrics.}
\label{realfirsttable}
\end{table}

A few recent papers recently propose countermeasures against LFI considering the bit flip model (PATRON, SPARSE)~\cite{choudhury2021patron,choudhury2021sparse} and the set-reset model (TAMED)~\cite{itcpaper}.
The latter incorporates the localization of the sensitive regions on the $DFF$ layout as assuming general bit flip model when devising countermeasures against LFI may result in a consequential error. To assess the vulnerability to LFI, PATRON proposes $VM$, SPARSE proposes $SVM$, and TAMED proposes $TVM$ as shown in the Table \ref{realfirsttable}. An additional utility of $VM$ is to measure if minimally discarding the less-than-adequate Hamming Distance (HD) encoding results in the necessary security as will be explained in Section~\ref{randd}. $TVM$ is a model-specific metric to capture FSM transition vulnerability whereas in the state-based approaches, $SVM$ and $VM$ protect \emph{all} $SS$ from the $NS$ equally according to the bit-flip model. The difference between $SVM$ and $VM$ is that only $SVM$ considers $FF$ layout for more precision. $TVM$ inherently assumes the FFs are a secure distance away from each other as it does not incorporate design layout information into the vulnerability estimation.  The reason, particularly $SVM$~\cite{choudhury2021sparse}
is chosen as a comparison metric because it comes closest in terms of vulnerability assessment by incorporating $\mathbb{FF}$ spatial information in the design, unlike another recently proposed metric ($TVM$) which although considers $\mathbb{AT}$ for bit flip and set-reset model, lacks in spatial information of the design~\cite{itcpaper}.


\begin{figure}[t]
  \centering
\subfloat[]{\includegraphics[clip,width=0.13\textwidth]{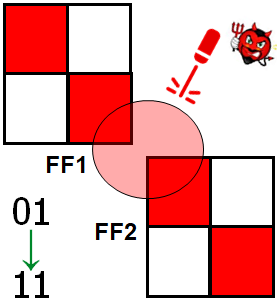}} \hspace{4mm}
\subfloat[]{\includegraphics[clip,width=0.13\textwidth]{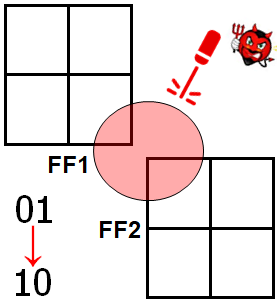}} \hspace{4mm} 
\subfloat[]{\includegraphics[clip,width=0.13\textwidth]{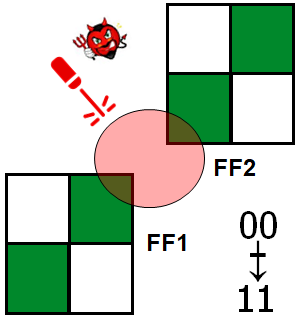}} 
  \caption{Comparison between set-reset and bit flip models under the same attack setting. Green and red arrows represent authorized and faulty transitions, respectively. Crossed arrow represents no transition occurs.
  (a) Set-reset model showing a laser incident on bit set sensitive regions of $FF_{1:2}$; (b) Bit flip model for the same attack setting as (a); (c) Set-reset model showing a laser incident on bit reset sensitive regions of $FF_{1:2}$.}
  \label{setresetbitflip}
\end{figure}

Figure~\ref{setresetbitflip} illustrates scenarios where the consideration of the precise set-reset model over the bit flip model becomes critical. In Fig.~\ref{setresetbitflip}(a) and (b), assuming the protected state set $\mathbb{P}=\{11\}$ and an authorized transition from state $\{01\}$ to $\mathbb{P}$, the set-reset model predicts the next state of $FF_{1:2}$ as $\{11\}$, which aligns with $\mathbb{P}$. However, the bit flip model predicts the next state as $\{10\}$, due to each state in $FF_{1:2}$ being flipped. If the FSM incorporates a security mechanism to detect transitions from $\mathbb{AT}$ to $\mathbb{P}$, it becomes evident that only the set-reset model can accurately identify $\mathbb{AT}$, whereas the bit flip model cannot. This may lead to a `false negative' scenario where a vulnerability might go unnoticed. Although the vulnerability assessment of set-reset model is more precise, there can be instances where unless all factors are considered, none of the previously proposed metrics provide appropriate security assessment.  

If Fig.~\ref{setresetbitflip}(a) is reconsidered where $FF_{1:2}$ are in close proximity, with $\mathbb{AT}=\{00 \rightarrow 11\}$,  ($\mathbb{AU}=\{00\}$,  $\mathbb{P}=\{11\}$), then as $HD(\mathbb{AU}, \mathbb{P})>1$, $VM=0$. $SVM>0$ assessment of the vulnerability in this case is correct since the laser is incident on the bit set sensitive regions of both FFs. However as the HD constraint is satisfied in TAMED, $TVM_{sr}=0$ would give the wrong vulnerability assessment. 

If Fig.~\ref{setresetbitflip}(c) is now considered with $\mathbb{AT}=\{00 \rightarrow 11\}$  ($\mathbb{AU}=\{00\}$, $\mathbb{P}=\{11\}$), then as $HD(\mathbb{AU}, \mathbb{P})>1$, $VM=0$. The current assessment of vulnerability ($SVM>0$) is inaccurate because the laser affects sensitive regions responsible for bit reset in both flip-flops. In the context of the bit-flip model, $SVM$ produces `false positives' by incorrectly identifying an $\mathbb{AT}$ event when no actual transition occurs. A correct vulnerability assessment ($TVM_{sr}=0$) would reflect the laser's impact on the reset-sensitive regions, aligning with the considered $\mathbb{AT}$. So for both the cases in (a) and (c) $VM$ cannot provide security at all as it does not consider spatial information of the FF and order of transition, $SVM$ provides `false positives' as it cannot realize the importance of the transition order considering only bit flip model, and $TVM$ lacks in spatial FF information in assessing the vulnerability so there's no way to correlate the correct encoding in $\mathbb{AT}$ with the corresponding FFs.

 \emph{In light of these examples, it is critical to incorporate a transitional approach which not only considers the order of transition, but also spatial inter-distance when identifying the FF vulnerability so that $TVM$ can be updated to assess all types of vulnerabilities comprehensively.} The proposed countermeasure must also possess a systematic approach for flexible sensitive area selection of the FFs, as previously mentioned. An updated metric known as spatial transitional vulnerability metric ($STVM$) is proposed in Section~\ref{stvm} to accomplish this.
This paper includes a security analysis of the bit-flip model, which can be induced alongside the more precise bit-set and bit-reset faults~\cite{dutertre2018laser}.

\section{TRANSPOSE Methodology} \label{TMe}

\subsection{Secure Flip-Flops and Normal Flip-Flops} \label{sffanffs}

Based on the preceding discussion emphasizing the spatial separation between specific FFs, we introduce two distinct FF categories within the SPARSE framework to constrain the spatial inter-distance between their sensitive regions in the TRANSPOSE framework. Designated as $\mathbb{SFF}$ and $\mathbb{NFF}$, these represent groups of \emph{secure FFs} and \emph{normal FFs}, respectively. The spatial separation of sensitive regions within each $\mathbb{SFF}$ group exceeds the spot diameter $D$ of the laser, preventing a single laser beam from affecting more than one $\mathbb{SFF}$ FF in a single clock cycle. Importantly, unlike in~\cite{choudhury2021sparse}, this definition of $\mathbb{SFF}$ is less restrictive in terms of area usage, focusing exclusively on specific groups within $\mathbb{FF}$. In $\mathbb{FF}$, FFs without spatial distance constraints between them are referred to as $\mathbb{NFF}$s. Depending on the spatial layout characteristics and the technology library, a laser spot could potentially affect one or more $\mathbb{NFF}$s within a single clock cycle.

The right side of Fig.~\ref{layoutsetresetexample} displays the $NFF$s and $SFF$s in light green and light red. As the state $\{1100\}$ can be overturned to access the $SS=\{0000\}$ according to the arrangement of the $\mathbb{NFF}$ in the layout, hence $ \{FF_1, FF_2\}$ is included as $\mathbb{NFF}$. These FF types are introduced to ensure that despite the potential for up to two single-bit flips per subset of $\mathbb{FF}$ with one laser each, the overall design layout of $\mathbb{FF}$ effectively counters LFI, enabling flexible area constraints to accommodate encoding requirements. As shown in Fig.~\ref{layoutsetresetexample}(a), although a maximum of 2 single bits of $11$ can be faulted to $00$ for the $\mathbb{SFF}$, this layout can still be used as a bit set model countermeasure as the set sensitive regions are spaced securely apart.
In this way, even if there are multiple lasers, $\mathbb{|SFF|}$ could be adjusted so that the HD in the $AT$ corresponding to the $\mathbb{SFF}$ bits is \emph{always kept higher} than the attacker's fault capability; the presence of $\mathbb{NFF}$ along with the less constrained definition of $\mathbb{SFF}$ than~\cite{choudhury2021sparse} facilitates area optimization in the design.

Thus, any given FSM design configuration can adjust $|\mathbb{SFF}|$, $|\mathbb{NFF}|$, $|\mathbb{AT}|$, and desired criteria $x$ to align with specific design needs, focusing on the few critical transitions that are pivotal in practice. Increasing $|\mathbb{SFF}|$ strategically can potentially allow for more permissible $\mathbb{AT}$, as discussed in Section~\ref{SPITATOT}. If augmenting $|\mathbb{SFF}|$ fails to meet security standards, then the number of FFs ($n$) must be increased accordingly.

\subsection{Formulating Spatial Transitional Vulnerability Metric} \label{stvm}

\begin{figure}[t]
  \centering
\subfloat[]{\includegraphics[clip,width=0.15\textwidth]{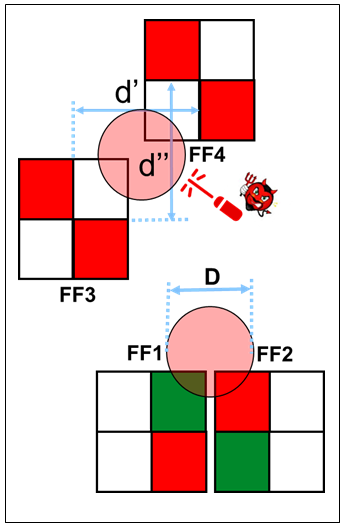}} 
\subfloat[]{\includegraphics[clip,width=0.15\textwidth]{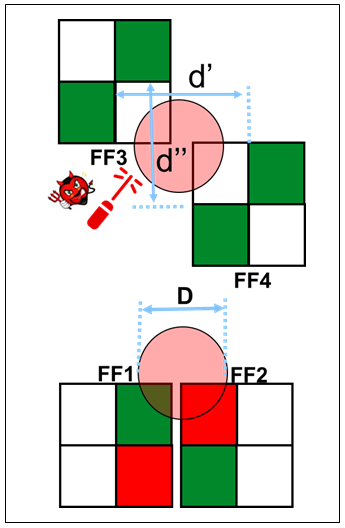}} 
\subfloat{\includegraphics[clip,width=0.20\textwidth]{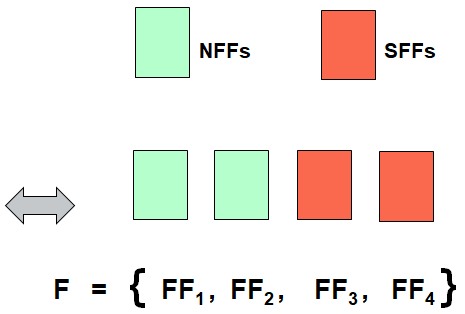}} 
  \caption{FF arrangement in an example design layout showing precise bit reset and set models under the attack setting of $x=1$ laser and beam diameter $D$. Normal and Secure FFs (NFFs and SFFs) are showm on the right. $d'$ and $d''$ represent the horizontal and vertical distances between the corresponding sensitive regions of $SFFs$; $d', d'' \leq D$. $FF_{1:2}$ is positioned in close proximity so $D$ can affect any combination of sensitive regions. LFI countermeasure corresponding to (a) bit reset model and (b) bit set model where a laser is incident on the bit reset and set sensitive regions of $FF_{3:4}$, respectively.
  }
  \label{layoutsetresetexample}
\end{figure}

In this section, we introduce the spatial transitional vulnerability metric ($STVM$). Unlike $VM$ or $SVM$, $STVM$ calculates the percentage of transitions where $x$ lasers can cause $f$ faults (where $f$ varies based on the number of FFs spatially affected), resulting in unauthorized transitions to a protected state ($P$). This calculation takes into account the ($y,z$) coordinates of the FFs in the physical layout. Since simultaneous alteration of the necessary number of FFs by $x$ laser spots may be constrained by spatial, temporal, or technical factors such as assumed fault models, there is merit in exploring a transitional approach. This approach leans towards a less conservative scenario, focusing on a subset of critical FFs from the total FFs ($n$) and their respective sensitive regions to unify relevant parameters. To this end, This approach integrates relative inter-distance information of FF-sensitive regions into vulnerability assessment, a factor overlooked by both $VM$ and $SVM$.

We start by introducing additional notation and an illustrative example. 
For a set of FFs in an FSM, $\mathbb{FF}=\{FF_1,\hdots, FF_n\}$ and we designate the laser location matrix $L=[\vec{l}_1, \vec{l}_2, \hdots, \vec{l}_x]$ where location $\vec{l}_i = [y_{i},z_{i}]$ represents the $y$ and $z$ coordinates of the $i$th laser.
The \emph{power set}, $\mathcal{P}(\mathbb{FF})$, is the set of all possible collections of FFs, expressed in sets. The goal is to define a set of FFs named as vulnerable FF set ($E_x$) that will represent the vulnerable FFs in a design based on the types of faults, number of lasers, relative inter-distances of the FF sensitive regions, and transitional information in the FSM. 

Here, we introduce the notations that model individual faults in specific FFs in order to denote vulnerability ($E_x$) precisely. If we are to assume the current state of a FF as $FF_y$ then, for the set-reset and bit-flip models the following notations (angle brackets and complements) can be used exclusively:
\begin{itemize}
  \item For bit set fault types, $FF_y \rangle =FF_y+1$. For bit reset fault types, $\langle FF_y=FF_y \cdot 0$. \textcolor{blue}{Hence, right and left angle brackets are used for bit set and bit reset, respectively}.
  \item For bit flip fault types, $\overline{FF_y}\; =\; \sim FF_y$.
\end{itemize}
$\langle{FF_y}\rangle$ denotes that according to the layout, the state of $FF_{y}$ can be changed with respect to the bit set and reset fault models, i.e., for bit set fault model with a given layout and $x$ laser, a current state of $0$ can be overturned to $1$ or a current state of $1$ stays at $1$ and vice versa for bit reset faults. Hence, these notations introduce the data dependent attributes of the set or reset fault models and the vulnerability of an individual or a combination of FFs in the layout can be captured as exemplified below.

To understand the precise vulnerability, a fixed FF layout for different positions of the laser beam should be precisely analyzed.
Fig.~\ref{layoutsetresetexample} shows an example layout of an FSM with arbitrary arrangement of $n=4$ FFs where different positions of the same laser ($x=1$) is depicted for clarity. To simplify, we assume the laser spot's power density is uniformly spread over its diameter, $D$. 
In Fig.~\ref{layoutsetresetexample}, where $x=1$ is assumed, $D$ is smaller than each of the FF horizontal and vertical inter-distances, $d',$ $d''$, meaning that a single laser is \emph{only} able to overturn the current state of the FF for $FF_3$, and $FF_4$, if their current state is `$1$' for (a) and `$0$' for (b). 
However, $FF_1$ is close enough to $FF_2$ such that one laser strike can affect any combinations of sensitive regions of $FF_{1:2}$. 
In this case, under the set-reset model, the \emph{combinations of vulnerable FF sets} according to the layout, $\mathbb{E}(x=1)=\{ \{\langle{FF_1}\rangle\}, \{\langle{FF_2}\rangle\}, \{\langle{FF_3}\rangle\}, \{\langle{FF_4}\rangle\}, \{\langle{FF_{1:2}}\rangle\}, \{\langle{FF_{3:4}}\}\}$ for (a) and $\{\{\langle{FF_1}\rangle\}, \{\langle{FF_2}\rangle\}, \{\langle{FF_3}\rangle\},  \{\langle{FF_4}\rangle\}$, $\{\langle{FF_{1:2}}\rangle\}$, $\{{FF_{3:4}}\rangle\}\}$ for (b).  
Each set in $\mathbb{E}(x)$ corresponds to a different position of the laser and the number of injected faults $f$ on specific FFs depends on the current state of the FF which is captured by these notations. For (a),  \{$\langle{FF_{3:4}}$\} denotes that this combination of FFs are only overturned in set-reset model, i.e. there is fault(s) only if at least one of the FF's current state is $1$ (if current state=$\{11\}$ then due to reset faults in both the FFs next state is $\{00\}$, and $f=2$). Note that the relative FF inter-distance with relation to $x$ lasers is automatically captured in $E_x$ based on the specific combinations of FFs, the faults can be effectuated on. Also, if the bit flip model was assumed, 
$\mathbb{E}(x=1)=\{ \{{\overline{FF_1}}\}, \{{\overline{FF_2}}\}, \{{\overline{FF_3}}\}, \{{\overline{FF_4}}\}, \{{\overline{FF_{1:2}}}\}, \{{\overline{FF_{3:4}}}\}\}$ for (a) and (b), i.e., there would be no differentiation and $E(x)$ would provide an inaccurate representation of FSM vulnerability.

In general, $\mathbb{E}(x) \subseteq \mathcal{P}(F)$ represents the set of vulnerable FF combinations invigorated with the possible types of fault model for any $L$. If all possible attack scenarios are represented by the function, $\beta \colon L \to \mathbb{E}(x)$, the set of FF fault types and combinations in one clock cycle is represented by $\{ \mathbb{V}(l_i) \;|\; l_i \in L\} \subseteq \mathbb{E}(x)$. In this way, the FF vulnerability according to the data dependent and data independent fault types can be represented precisely.

Using these definitions, we can provide an expression for $\mathbb{SVT}(x)$. 
The spatially vulnerable transition set for Fig.~\ref{layoutsetresetexample}(a) is $\{XX11 \rightarrow XX00, XX10 \rightarrow XX00, XX01 \rightarrow XX00, XX11 \rightarrow XX01, XX11 \rightarrow XX10\}$, where the first transition is included because  although the $HD=2$ requirement is met, $f=2$ according to the layout and $AU$, and for the subsequent transitions the $HD$ constraint is not met in the $SFF$ bits. In the same way, spatially vulnerable transition set for Fig.~\ref{layoutsetresetexample}(b) is $\{XX00 \rightarrow XX11, XX00 \rightarrow XX10, XX00 \rightarrow XX01, XX01 \rightarrow XX11, XX10 \rightarrow XX11\}$. 
The spatial vulnerable transition set can be easily inferred from $\mathbb{E}(x)$. Hence,  $\mathbb{SVT}(x)$ is the spatially vulnerable transition set susceptible to $x$ lasers~\emph{considering the state FF layout} and $\mathbb{AT}$ in the design. 
 
The FSM's degree of susceptibility to $x$ laser based faults in one clock cycle based on the $\mathbb{AT}$ is captured by Spatially Transitional Vulnerability Metric, $STVM(x)$,
\begin{equation}
STVM (x)= \dfrac {   \big| [ \bigcup\limits_{A \in \mathbb{V}(l_i)} A ]   \subseteq \mathbb{FF}    \big|  } {   |\mathbb{T}|}
\end{equation}
The numerator of $STVM(x)$ represents the set of vulnerable FF fault
types and combinations in one clock cycle according to the FF layout in the design for a precise laser location, $(l_i)$, which could be extrapolated to the \emph{spatially vulnerable transitions set}.
Note that this is a general expression for $STVM$ and can be extended to the bit-flip ($STVM_{bf}$), and set-reset ($STVM_{sr}$) models by choosing to represent the $\mathbb{SVT}$ with either of the bit-flip or set-reset models notations i.e., complements and angle brackets, respectively.  
The desired value of spatially vulnerable transitions set, $\mathbb{SVT}(x)=\emptyset$. Note that, $STVM_{bf}(x)>VM(x)$ or $STVM_{sr}(x)>VM(x)$ signifies that at least one pair of FFs can be flipped with one laser spot in one clock cycle (i.e., $f>x$) for at least one set of laser coordinates. Intuitively, $STVM_{bf}(x)>(STVM_{sr}(x)=0)$ signifies that the current $\mathbb{AT}$ considered according to the $\mathbb{FF}$ layout and laser position, $(l_i)$ is secure according to the set-reset model and showing a false positive for the bit-flip model.

\subsection{Salient Parameters and Transition Types in TRANSPOSE} \label{SPITATOT}

\begin{figure}[t]
  \centering
  \includegraphics[width=0.9\linewidth]{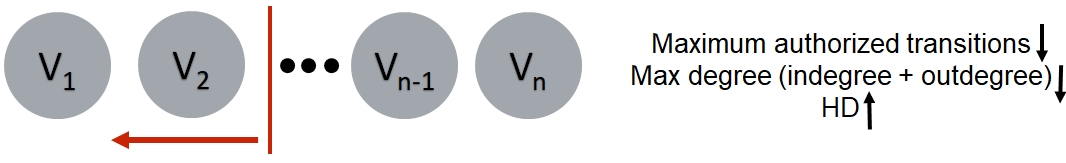}
  \caption{Relationship between salient parameters of TRANSPOSE with the increase of security bits/$\mathbb{SFF}$.}
  \label{longarrow}
\end{figure}

\noindent \textbf{Parameters:} From Section~\ref{FSMencoding}, an FSM is represented by vertices and edges. Let $\mathbb{S}$ denote a finite set of states, where each state in $\mathbb{S}$ is represented by a vector of length $n$: $[v_{1},v_{2}, \hdots v_{n}]$, $v_{i} \in \{ 0, 1\} \forall i$. Here, $v_{i}$ represents the variable associated with the $i^{th}$ FF in the FSM.
As a convention (without loss of generality), we assume that $m=|\mathbb{SFF}|$ rightmost bit positions of the vectors correspond to $\mathbb{SFF}$ unless specified otherwise. For instance, the $\mathbb{SFF}$ correspond to $[v_{n-m+1},\hdots, v_{n}]$ (abbreviated as $\vec{v}_{n-m+1:n}$) for states in $\mathbb{S}$. These variables are illustrated in Fig.~\ref{longarrow}  along with various pertinent parameters for optimizing encoding within the TRANSPOSE methodology. 

The vertical red demarcation line delineates $\mathbb{SFF}$ from $\mathbb{NFF}$. According to our convention, the requirement of achieving $HD \geq x$ is applied to $\mathbb{SFF}$. On the left-hand side, unless specified otherwise, we assume don't care bits (denoted as X), which are represented by $\mathbb{NFF}$. In our effort to minimize dynamic power consumption through fault injection-resistant FSM design, it is crucial to incorporate as many don't care bits (Xs) as possible. This approach maximizes flexibility for optimizing switching activity in state encoding. It is also desirable to implement $\mathbb{NFF}$ because of no spatial constraint as explained in Section~\ref{sffanffs}. TRANSPOSE is constructed so that maximum possible number of Xs is selected and optimized according to the FSM switching activity, keeping in mind that more $\mathbb{SFF}$ means more area constraints in the design. 

As the demarcation line shifts leftward, the potential number of authorized transitions within the FSM intuitively diminishes. This reduction is visualized as fewer Xs limit the encoding flexibility to variations of `$0$' or `$1$'. For instance, the achievable $|\mathbb{AT}|$ in '$XX00 \rightarrow XX11$' exceeds that in '$X000 \rightarrow X111$' when LFI capability $x= 1$. Furthermore, increasing $|\mathbb{SFF}|$ (\emph{security bits}) typically enhances the HD capability due to a greater number of bits dedicated to $\mathbb{SFF}$.
The parameters of indegree and outdegree are critical, potentially necessitating a higher $n$ to meet HD requirements within $\mathbb{AT}$. Indegree refers to the number of incoming edges to a vertex, while outdegree denotes the number of outgoing edges. Increasing the number of Xs naturally expands the capacity to accommodate higher indegree and outdegree configurations for vertices, as illustrated in specific examples below.

\begin{figure}[t] 
\centering
\subfloat[]{\includegraphics[clip,width=0.15\textwidth]{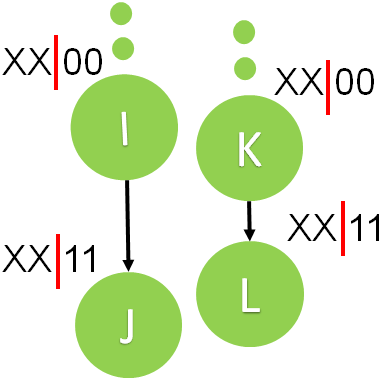}} \hspace{0.05em}%
\subfloat[]{\includegraphics[clip,width=0.15\textwidth]{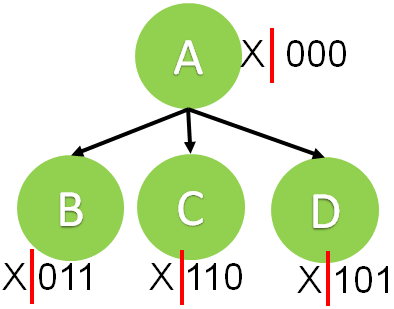}}  \hspace{0.05em}%
\subfloat[]{\includegraphics[clip,width=0.15\textwidth]{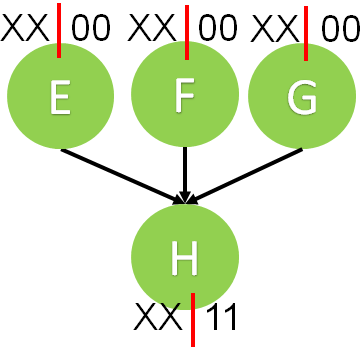}}   \hspace{0.05em}%
\subfloat[]{\includegraphics[clip,width=0.19\textwidth]{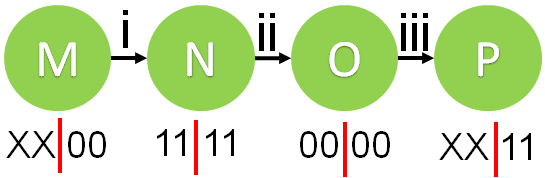}}
\caption{TRANSPOSE constructs a subset of $\mathbb{AT}$ variations, all assuming an LFI capability of $x=1$. In these examples, X denotes don't care states in the encoding. The red demarcation line in figures (a)-(c) distinguishes don't care bits ($\mathbb{NFF}$) from security bits ($\mathbb{SFF}$).  All bits not designated as X are considered security bits. (a) $I \rightarrow J,K \rightarrow L$ occur within the same FSM. (b)(c)(d) each depicts 3 protected transitions.}
\label{Ttypes}
\end{figure}

\begin{figure*}[t]
  \centering
  \includegraphics[width=0.98\linewidth]{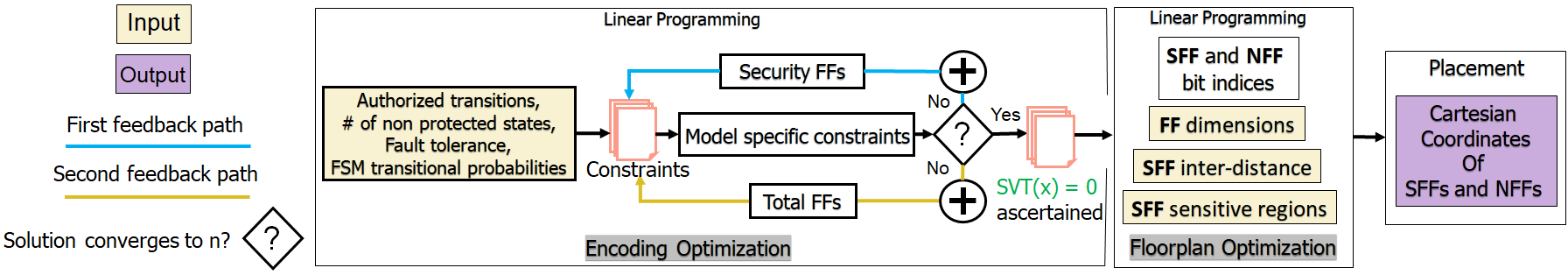}
  \caption{Block diagram of TRANSPOSE.}
  \label{tamedframeowrk}
\end{figure*}

\noindent \textbf{Transition types and examples:} Fig.~\ref{Ttypes} depicts a subset of $\mathbb{AT}$ and corresponding TRANSPOSE state encodings. In all these examples, $x=1$ is assumed, ensuring a minimum $HD=2$ between transitions secured by the spatial inter-distance of $\mathbb{SFF}$ sensitive regions. This guarantees security across each transition pair. The examples are defined in accordance with the tree-based data structure. A useful term, \emph{depth} of a vertex is defined as the number of edges (transitions) in the path from \emph{root} to the vertex, where the root is simply the node of reference.

\textbf{\textit{No. of origin roots = No. of ending nodes}}  In this case, the maximum depth for all the $\mathbb{AT}=1$, and the total $|\mathbb{AT}|=$ ({\tiny $\dfrac{total \: number \: of \: nodes \: involved} {2}$}). The two transitions depicted in Figure~\ref{Ttypes}(a) belong to the same FSM. It is evident that there is potential for optimization using Xs with $n=4$ and a consistent $HD=2$ within the security bits ($\mathbb{SFF}$).

\textbf{\textit{No. of origin root (= 1) $<$ No. of ending nodes}} 
Although, the maximum depth in this case is still at 1, there can be many edges originating from the same root node requiring protection.
Moving the demarcation line for the outdegree of node $A$ in Figure~\ref{Ttypes}(b) results in fewer Xs within the security bits ($\mathbb{SFF}$) for the same $HD$. Consequently, this reduces the number of possible combinations for authorized transitions. For instance, $0000 \rightarrow \{0011, 0110, 0101\}$ exemplifies this scenario.

\textbf{\textit{No. of origin roots $>$ No. of ending node (= 1)}} Just like the previous example, there can be many originating nodes transitioning to the same ending node requiring protection.  Repositioning the line to its original position is essential for the indegree of node $H$ in Fig.~\ref{Ttypes}(c). This adjustment meets the $HD$ requirement within $\mathbb{SFF}$ and ensures sufficient unique combinations among nodes (represented by X/$\mathbb{NFF}$), namely $E$, $F$, and $G$. For example, $\{0000, 0100, 1000\} \rightarrow 0011$ exemplifies this condition.

\textbf{\textit{No. of origin roots = No. of ending nodes}} 
 In this case, the maximum depth considering all the $\mathbb{AT}$ path $>$ 1 and the total $|\mathbb{AT}|=$ (total number of nodes involved -1). Fig.~\ref{Ttypes}(d) presents an illustration where a sequence of consecutive authorized transitions ($i$, $ii$, and $iii$) originates from the same terminal node.
 
  In this paper, we categorize this type of FSM as a \emph{directed rooted tree FSM}. Here, the security bits ($\mathbb{SFF}$) extend to include the leftmost two bits due to the requirement of a $1 \rightarrow 0$ transition in the $\mathbb{AT}$ ($ii$). Despite a further reduction in don't cares (no $\mathbb{NFF}$ in this example), $n$ remains at 4 to conserve area and power. In this scenario, the left-hand bits adhere to the bit set model under the set-reset, securing the $\mathbb{SFF}$'s reset-sensitive regions by the $1' \rightarrow 0'$ transition. Conversely, the security bits on the right-hand side adhere to the conventional bit reset model, securing the $\mathbb{SFF}$'s set-sensitive regions. As the security bits incorporate both types of transitions for security we refer to this methodology as set-reset approach under the set-reset model. TRANSPOSE can also incorporate set only (only $1 \rightarrow 0$ transition) and reset only (only $0 \rightarrow 1$ transition) under the set-reset model, with spatial security applied solely to the reset-sensitive or set-sensitive regions of $\mathbb{SFF}$, respectively. 
 
 In this paper, results and analysis of all the approaches under set-reset model have been included as all of these approaches can provide appropriate security.

\subsection{TRANSPOSE Framework}  


In this section, we introduce TRANSPOSE (\underline{TRANS}itional A\underline{P}proaches f\underline{O}r \underline{S}patially-Aware LFI Resilient State Machine \underline{E}ncoding), our automated framework for encoding and placement. It aims to enhance FSM resilience against precise laser fault injection using transition-based models. We employ Integer Linear Programming (ILP) to achieve optimized encoding that minimizes switching activity and ensures security with $|\mathbb{SVT}|=0$, aligned with specified design requirements and user inputs. TRANSPOSE comprises two main components: \emph{secure encoding optimization} and \emph{floorplan optimization}, as depicted in the block diagram in Figure~\ref{tamedframeowrk}. States not involved in authorized transitions are termed `Non-protected states' ($\mathbb{NS}$), detailed in Section~\ref{FSMencoding}.

The user inputs required for TRANSPOSE include the \emph{design specification} (FSM transitions, $\mathbb{T}$, and corresponding transitional probabilities for optimizing switching activity), \emph{FSM security requirements} (such as authorized transitions, number of non-protected states), and the \emph{capabilities of the attacker} (e.g., number of laser faults $x$). Additionally, inputs like $\mathbb{FF}$ physical dimensions, expected inter-distance between $\mathbb{SFF}$, and sensitive regions of $\mathbb{SFF}$ are necessary. The choice of transitional approach (bit flip or set-reset as discussed in this paper) can also be specified as supplementary input.

All transitional information $\mathbb{T}$ is inputted using an adjacency list where the default transition order is considered significant. Once the inputs are provided, the corresponding linear constraints for $n = \log_2 |\mathbb{S}|$ FFs are initialized. Subsequently, an iterative process begins to determine if a solution converges with the current $n$ FFs. During each iteration, $\mathbb{|SFF|}$ is initially increased (denoted by the blue line), and the associated linear constraints are updated accordingly to verify compliance with the design specifications. If the specifications are not met, $n$ is incremented by 1, and the second feedback path (yellow line) is explored. This iterative process continues until the ILP process converges to an appropriate $n$, and an optimized state encoding is generated by TRANSPOSE. Finally, the TRANSPOSE encoding is validated to ensure $SVT_x = 0$, depending on the fault model, thereby preventing any authorized transition $AT$ from failing to meet the security constraint.

In the floorplan optimization block, information regarding $\mathbb{SFF}$ and $\mathbb{NFF}$ indices is passed from the preceding block. Additional inputs include the dimensions of FFs (determined by the technology node/process design kit) and the minimum distance (defined by the laser spot size). Subsequently, integer linear programming (ILP) is employed to compute the Cartesian coordinates of all FFs in the FSM layout. TRANSPOSE ensures that the $\mathbb{SFF}$ are strategically spaced apart to secure their sensitive regions, as required by the security bits in $\mathbb{AT}$, while minimizing the area used, facilitated by the placement of $\mathbb{NFF}$. Note that, contrary to SPARSE~\cite{choudhury2021sparse}, TRANSPOSE does not necessitate all $\mathbb{NFF}$ to also be placed a secure distance away from $\mathbb{SFF}$. The internal procedures of encoding and floorplan optimization are elaborated in Sections~\ref{SEO} and~\ref{fo}, respectively.

\subsubsection{Secure Encoding Optimization} \label{SEO}   

\noindent Our objective is to develop a framework capable of integrating various transitional approach models (bit flip, reset only, set only, set and reset) with appropriate ILP constraints to ensure an LFI-resistant FSM. The objective function focuses on minimizing the total switching activity of the FSM to optimize dynamic power consumption. Below, we provide detailed explanations of the common and distinct constraints, as well as the objective function employed in the framework. \textcolor{blue}{ILP is NP-Complete}. 

For each state encoding $r_{i}$, where $i=1, \hdots, g$, of an $g$-state FSM, our objective function for the linear optimization problem can be expressed as finding a code, $[r_{i,1}, r_{i,2}, \hdots r_{i,n}]$, such that:
\begin{equation}
\begin{split}
& \min_r h(r) \; \textrm{where} \\
& h(r) =   \sum \limits_{1 \leq i < j \leq g} p_{i,j} \sum \limits^n_{l=1} \left| r_{il}-r_{jl} \right|,  i \neq j \\
& \forall r_{il} \in \{ \: g - state \: encoding \:  \}, \\
& subject \, to \, \left\{\begin{split}
& Constraints \\
& r \: ( 0\:-1)\: integer
\end{split}\right.
\end{split} 
\end{equation}
where $n$ is the number of FFs in the FSM design; $p_{i,j}$ represents the total transitional probability between states $r_{il}$ and $r_{jl}$, where $l$ represents the index of bits in the state encoding. Here, the dimension $i$ corresponds to each of the state FFs. The ILP constraints for each of the model are elaborated upon below. 

\noindent{\bfseries Bit-Flip Model:} The initial constraint ensures compliance with the design specifications for authorized transitions ($\mathbb{AT}$) between the sets $\mathbb{AU}$ and $\mathbb{P}$.  Considering the attacker's LFI capability of $x$, all states $\in \mathbb{AU}$ must maintain a minimum HD of  $x+1$ from all states $\in \mathbb{P}$. This is expressed as:
	\begin{equation}
	\label{eq:14}
	 \sum\limits^n_{l=1} \left| r_{AU l} - r_{Pl} \right| > x
	\end{equation}

Assuming no self-transitions, the total number of possible combinations of transitions in an FSM can be calculated using the combination function as $|\mathbb{S}|C_2$ in an FSM. Here, $C$ denotes the combination function. All combinations of transitions excluding those in $\mathbb{AT}$, i.e., ($\mathbb{|S|}C_2 - |\mathbb{AT}|$) must be at least unit $HD$ away. This requirement applies to combinations of transitions that do not exist in $\mathbb{T}$ within the FSM, expressed as $\sum\limits^n_{l=1} \left| r_{AU_al} - r_{AU_bl} \right| \geq 1$,  $\sum\limits^n_{l=1} \left| r_{P_al} - r_{P_bl} \right| \geq 1$, $\sum\limits^n_{l=1} \left| r_{NS_al} - r_{NS_bl} \right| \geq 1$, $\sum\limits^n_{l=1} \left| r_{NSl} - r_{AUl} \right| \geq 1$, and $\sum\limits^n_{l=1} \left| r_{NSl} - r_{Pl} \right| \geq 1$, where $a \neq b$.
Thus, this constraint ensures that each state within $\mathbb{T}$ is distinct in the FSM.

\noindent{\bfseries Set-Reset Model:}
To clarify, in the set-reset model, we adopt a convention where the rightmost security bits follow the reset model by default, and if dictated by $\mathbb{AT}$, the leftmost security bits adhere to the set model. In addition to the standard constraints of the bit flip model, specific constraints unique to the set-reset model are necessary, as outlined below.

Adhering to this convention, each iteration may necessitate certain security bits to transition specifically from $0$ to $1$ bits to meet security requirements, while striving to achieve the optimal value of $n$.  
    The constraints specified in Equations~\eqref{eq20} and~\eqref{eq21} serve this purpose. Given an attacker's capability $x$ in LFI, it's essential that at least ($x+1$) `0' bits are initially required in the rightmost ($m=|\mathbb{SFF}|$) security bits of $AU$
    This requirement is expressed as:   
	\begin{equation}
	\label{eq20}
	 \sum\limits^m_{l=1}  r_{AUl} \leq  m-(x+1)
	\end{equation}
Moreover, it's crucial that if a security bit in $AU$
is `0' the corresponding bit in $P$ must be `1' among the $m$ security bits. However, if the bit in $AU$ is `1', the corresponding bit in $P$ could be either `0' or `1'. This condition is captured by: 
    \begin{equation}
	\label{eq21}
	  r_{Pl} \geq  1-r_{AUl},\; \textrm{where} \; \forall l=1,\dots,m
	\end{equation}
Alternatively, if there is a requirement for a \( 1 \rightarrow 0 \) to adhere to the bit set model in the leftmost security bits for FSMs similar to Fig.~\ref{Ttypes}(d), the aforementioned constraints are modified as:
	\begin{align}
	\label{eq22}
	 \sum\limits^m_{l=1} 1- mr_{AUl} \leq  m-(x+1) \\
	 \label{eq25}
	  r_{Pl} \leq  1-r_{AUl},\; \textrm{where} \; \forall l=1,\dots,m
	\end{align}	
Note that, the above constraints implemented separately and together lead to the genesis of all the set-reset models (set only, reset only, set and reset). For example, to implement the `set only' approach, equations~\eqref{eq22}-\eqref{eq25} can be used in both righmost and leftmost security bits, if needed.

\subsubsection{Floorplan Optimization} \label{fo}   
\begin{figure}[]
  \centering
  \includegraphics[width=0.9\linewidth]{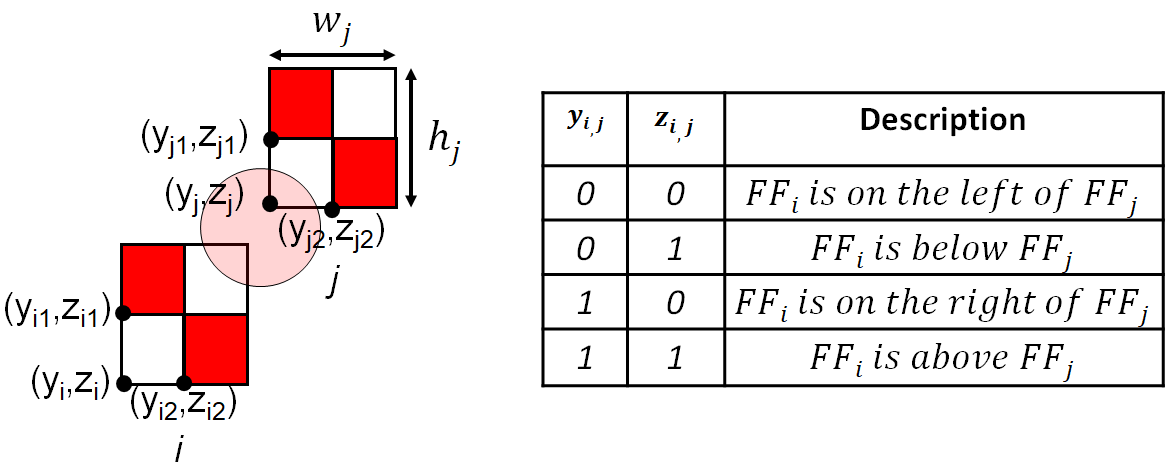}
  \caption{Layout of two $\mathbb{SFF}$ with precise co-ordinates (unknown integer variables) realizing sensitive regions of set-reset model; positional notations of the unknown binary variables used in ILP constraints are described in the adjacent table.}
  \label{securityplacement}
\end{figure}

Floorplan optimization determines the optimal placement of $\mathbb{SFF}$ and $\mathbb{NFF}$ with minimal area. For brevity, uniform shapes for all FFs, specifically fixed width ($w_i$) and height ($h_i$) for the \textit{i}th FF ($FF_i$) among $n$ total FFs are assumed. The integer variables, $y_i$ and $z_i$, denote the coordinates of the lower-left vertex of $FF_i$. The binary variables, $y_{ij}$, $z_{ij} \in \{0,1\}$  represent the relative positional information between FFs $i$ and $j$ as illustrated in the descriptive table in Fig.~\ref{securityplacement}. $H$ and $W$ denote the upper bounds for the floorplan height and width. Due to the inherent nonlinearity of area minimization (width $\times$ height), our approach optimizes an objective function minimizing the floorplan's height (denoted as $Y$), assuming an initial width.
The linear constraints are:
\begin{align}
y_i + w_i \leq W, \;\;\; &1 \leq i \leq n  \label{eq:fo1}\\
z_i + h_i \leq Y, \;\;\;1 &\leq i \leq n  \label{eq:fo2}\\
y_i,z_i \geq 0, \;\;\;1 &\leq i \leq n \label{eq:f7}
 \end{align}
Equations~\eqref{eq:fo1} and \eqref{eq:fo2} ensure that each FF is contained within the defined floorplan limits, while Equation~\eqref{eq:f7} ensures that the coordinates of $\mathbb{FF}$ are restricted to non-negative integers. Following constraints ensure that none of the $\mathbb{FF}$ overlap:
\begin{align}
&z_i + w_i \leq z_j + W (y_{ij}+z_{ij}) ,\;\;   i \neq j \label{eq:fo3}\\
&z_i - w_j \geq z_j - W (1- y_{ij}+z_{ij}),\;\;   i \neq j \label{eq:fo4}\\
&y_i + h_i \leq y_j + H (1+ y_{ij}-z_{ij}),\;\;   i \neq j \label{eq:fo5}\\
&y_i - h_j \geq y_j - H (2- y_{ij}-z_{ij}), \;\;   i \neq j  \label{eq:fo6} 
\end{align}
where $FF_{i,j} \in \mathbb{FF}$, $i \neq j$. The security constraints can be understood by the two $\mathbb{SFF}$ examples shown in Fig.~\ref{securityplacement}. First, the equality constraints to realize $\mathbb{SFF}$ sensitive regions are 
\begin{align}
&y_i = y_{i1},\; y_i = y_{i2}- (w_i/2) \\
&y_j = y_{j1},\; y_j = y_{j2}- (w_j/2)  \\
&z_i = z_{i2},\; z_i = z_{i1}- (h_i/2) \\ 
&z_j = z_{j2},\; z_j = z_{j1}- (h_j/2)
\end{align}
And secondly, the security constraints are
\begin{align}
&y_{j2} - y_{i2} \geq D \label{eq:lauuut}\\
&z_{j1} - z_{i1} \geq D \label{eq:last} 
\end{align}
Note that these equality and security constraints can be varied and extended to other $\mathbb{SFF}$ of different sensitive regions according to how the designer wants. The constant $D$ denotes the diameter of the laser beam, representing a secure spatial separation between sensitive regions within $\mathbb{SFF}$, as depicted in Equations~\eqref{eq:lauuut}-\eqref{eq:last}. For the bit flip model, only the Equations~\eqref{eq:fo1}-\eqref{eq:fo6} are used. As in bit flip model, all $\mathbb{FF}$ are $\mathbb{SFF}$ a variation of equations~\eqref{eq:fo3}-\eqref{eq:fo6} are used to incorporate $D$~\cite{choudhury2021sparse}. Equations~\eqref{eq:fo1}-\eqref{eq:last} are used in set-reset (set only, reset only, and set and reset) model. 

To minimize the floorplan area,  one can iterate over different widths $W$ and solve the ILP problem iteratively, selecting the configuration that yields the smallest area ($W \times Y$). The resulting Cartesian coordinates of the FFs in $\mathbb{SFF}$ and $\mathbb{NFF}$, along with the secure encoding generated from the encoding optimization phase, guarantee $STVM(x) = 0$, ensuring the FSM's resilience against LFI.

\section{Results and Discussion} \label{randd}

In this section, we assess the proposed TRANSPOSE encoding and contrast it with alternative FF encoding schemes. Our evaluation focuses on the post-synthesis outcomes, specifically: (i) the Power Delay Product (PDP) of the entire design normalized by binary encoded FSM, (ii) the power consumption of the FSM encoded module normalized by the binary encoded FSM, and (iii) the area of the overall design normalized by the binary encoded FSM area. Although the correlation of PDPs with area and power is well understood, individual values for each $x$ are provided to assess the cumulative impact of local and global changes induced by TRANSPOSE encoding, in conjunction with other considerations such as manually selected $SS$ for PATRON and SPARSE, randomized state allocations for Codetables, and tool optimizations. 

The following security metrics are also compared: $VM$, $STVM_{bf/sr}$, and $SVM$ with increasing $x$. In our assessment, the laser spot diameter $D$ chosen for calculating spatially-aware security metrics is set to 1$\mu m$~\cite{boit2016ic}. We maintain proximity to current technological norms for simplicity and consistency. Note that, although the hardness of LFI varies with the targeted geometry size, the effective laser beam diameter, and other physical as well as device and laser inherent functionalities, the consequent effects, and the manifestation of vulnerabilities can be rationalized with the same physical explanations~\cite{richter2022revisiting}. Hence, a different $D$ value may also manifest a similar vulnerability demonstrated by the experiments. Furthermore, the minimum achievable laser spot diameter has not yet been successfully reduced below $1\mu m$ due to optical diffraction limitations~\cite{agoyan2010single}. Regardless, the most precise $D$ is considered in our experiments as the vulnerability manifestations of the variations of $D$ would still have the same physical principles. Finally, while current constraints typically limit the number of simultaneous laser faults to two (i.e., $x \leq 2$) today,  we also present results for $x=3$ to illustrate the robustness of our threat model and the framework's readiness to address future LFI attack scenarios, as discussed in~\cite{bossuet2021multi}. 

 \vspace{0.5ex}

\noindent \textbf{Benchmark FSMs:} TRANSPOSE and the other encoding schemes are investigated on five controller benchmark circuits, namely AES, SHA, FSM Controller, Power Sequencer and Versatile IIR Filter (VIIRF). Some of these specific benchmarks are different than the previous papers~\cite{choudhury2021patron ,choudhury2021sparse, itcpaper} due to the unavailability of all the modules in benchmarks at the time of writing this paper~\cite{opencore}. Note that, the whole design is required to consider the transitional probabilities accurately. Regardless, important benchmarks such as popular encryption engines, vendor independent and stable constructs of a configurable IIR filter, power supply sequencer design incorporating the capability of handling multiple supply voltages in large electronics systems are specifically chosen as these are components relatable in industry use.
All benchmark circuits originate from from OpenCores~\cite{opencore} except for the synthetic benchmark named `FSM Controller' and synthesized using Synopsys Design Compiler (DC) with 32-nm library. \textcolor{blue}{To enforce proper spatial distances among FFs, IC Compiler II (ICC2) is automated using the $create\_rp\_group$ command}. Cartesian coordinates of $\mathbb{SFF}$ and $\mathbb{NFF}$ are derived using the $get\_attribute$ command and fed into an in-house tool for analysis, which evaluates the design at laser positions $(l_i)$ across the entire layout (at intervals of 0.1$\mu m$ along $y$ and $z$ axes), computing $STVM(x)$ and $SVM(x)$ metrics accordingly. 
\textcolor{blue}{All the overhead calculations are performed post removal of FSM optimization pass during synthesis. To maintain FSM security and to ensure FSM encoding remains unaltered the command $set\_fsm\_encoding$ is used}.

\begin{table}[]
\centering
\scalebox{1}
{%
\begin{tabular}{c|ccccc|}
\cline{2-6}
\multicolumn{1}{l|}{}    & \multicolumn{1}{c|}{AES} & \multicolumn{1}{c|}{SHA-256} & \multicolumn{1}{c|}{\makecell{FSM \\ Controller}} & \multicolumn{1}{c|}{\makecell{Power \\ Sequencer}} & \multicolumn{1}{c|}{VIIRF} \\ \hline
\multicolumn{1}{|c|}{$\mathbb{|S|}$}  & 5                        & 7                            & 7                        & 9                                   &            12                                 \\ \cline{1-1}
\multicolumn{1}{|c|}{$\mathbb{|T|}$}  & 10                       & 11                           & 9                        & 11                                   & 13                                            \\ \cline{1-1}
\multicolumn{1}{|c|}{$\mathbb{|SS|}$} & 3                        & 3                            & 4                        & 4                                   & 6                                            \\ \cline{1-1}
\multicolumn{1}{|c|}{$\mathbb{|AT|}$} & 2                        & 2                            & 2                        & 3                                   & 3                                            \\ \cline{1-6}
\end{tabular}%
}
\caption{The associated metrics include the total number of states ($\mathbb{|S|}$), total number of transitions ($\mathbb{|T|}$), total number of sensitive states ($\mathbb{|SS|}$), and total number of authorized transitions ($\mathbb{|AT|}$) for each benchmark.}
\label{firsttable}
\end{table}

\vspace{0.5ex}

\noindent \textbf{FSM Encoding Schemes:} TRANSPOSE is compared with PATRON and SPARSE that can \emph{only} assume the bit flip model~\cite{choudhury2021patron, choudhury2021sparse}. As PATRON and SPARSE schemes cannot generate a \emph{singular} power optimized and FSM transitional probabilities incorporated encoding, the process of obtaining one optimum encoding can be manually exhaustive; multiple encodings meeting the same FSM design constraint are possible. Therefore, the overhead metrics for PATRON and SPARSE encompass the average of five distinct values for each $x$. 

The linear codes, termed as Codetables in this paper~\cite{Grassl:Codetables} are also compared with TRANSPOSE due to their established effectiveness against LFI. An $[n,k,d]$ linear code comprises $k$-bit messages within $n$-bit codewords, where any two distinct codewords differ by at least $d$ bits. Codetables detail the boundaries and construction of such linear codes over the Galois Field of order $q$. Since we focus solely on Boolean values here, $q=2$, hence only binary codes are considered. In the case of Codetables, all transitions $\mathbb{T}$ in the FSM are regarded as authorized transitions $\mathbb{AT}$, without the flexibility to \emph{distinguish between $AT$ and $(T-AT)$ transitions}. Codetables encompass various popular linear codes such as Hamming (7, 4), Extended Hamming, Binary Golay, Extended Binary Golay, etc., by design. The uniform Hamming distance ($x+1$) between codewords is assumed, and the average of five different values is computed for Codetables as well.

Most cryptographic algorithms typically exhibit a limited number of states. For instance, in AES, the authorized transitions encompass movements from the ``Initial Round'' to ``Do Round'' and ``Do Round'' to ``Final Round''. Similarly, in SHA-256, transitions from ``Block next'' to ``Data input'', and ``Valid'' are recognized as authorized transitions. 
For each benchmark, $\mathbb{AT}$ is selected to encompass all distinct transition types as illustrated in Fig.~\ref{Ttypes} ensuring a comprehensive evaluation of the potential quantitative cost and qualitative security to check for adopting TRANSPOSE. Compared to TRANSPOSE, PATRON~\cite{choudhury2021patron} and Codetables~\cite{Grassl:Codetables} consider states for the solution set(s) instead of the $AT$. 
Moreover, TRANSPOSE offers flexibility to interpret all $\mathbb{T}$ transitions within the FSM as $\mathbb{AT}$ if desired by the designer.
\begin{table*}[]
\centering
\small
\scalebox{0.8}
{\begin{tabular}{llllllllllllllllllll}
\cline{3-17}
\multicolumn{1}{c}{}                                          & \multicolumn{1}{c|}{}      & \multicolumn{3}{c|}{AES}                                                       & \multicolumn{3}{c|}{SHA-256}                                                   & \multicolumn{3}{c|}{FSM   Controller}                                           & \multicolumn{3}{c|}{Power  Sequencer}                                          & \multicolumn{3}{c|}{VIIRF}                                                     &  &  &  \\ \cline{2-17}
\multicolumn{1}{c|}{}                                         & \multicolumn{1}{c|}{x}     & \multicolumn{1}{c|}{1}   & \multicolumn{1}{c|}{2}   & \multicolumn{1}{c|}{3}   & \multicolumn{1}{c|}{1}   & \multicolumn{1}{c|}{2}   & \multicolumn{1}{c|}{3}   & \multicolumn{1}{c|}{1}    & \multicolumn{1}{c|}{2}   & \multicolumn{1}{c|}{3}   & \multicolumn{1}{c|}{1}   & \multicolumn{1}{c|}{2}   & \multicolumn{1}{c|}{3}   & \multicolumn{1}{c|}{1}   & \multicolumn{1}{c|}{2}   & \multicolumn{1}{c|}{3}   &  &  &  \\ \cline{1-17}
\multicolumn{1}{|c|}{\multirow{3}{*}{ TRANSPOSE (Bit-Flip)}}      & \multicolumn{1}{c|}{PDP}   & \multicolumn{1}{c}{1.01} & \multicolumn{1}{c}{1.03} & \multicolumn{1}{c}{1.02} & \multicolumn{1}{c}{0.98} & \multicolumn{1}{c}{1.05} & \multicolumn{1}{c}{1.09} & \multicolumn{1}{c}{\textbf{1.01}}  & \multicolumn{1}{c}{\textbf{1.01}} & \multicolumn{1}{c}{\textbf{1.2}}  & \multicolumn{1}{c}{1.12} & \multicolumn{1}{c}{\textbf{1.17}} & \multicolumn{1}{c}{\textbf{1.24}} & \multicolumn{1}{c}{\textbf{0.98}} & \multicolumn{1}{c}{\textbf{0.99}} & \multicolumn{1}{c|}{\textbf{1.01}} &  &  &  \\ \cline{2-2}
\multicolumn{1}{|c|}{}                                        & \multicolumn{1}{c|}{Power} & \multicolumn{1}{c}{\textbf{0.94}} & \multicolumn{1}{c}{\textbf{1.12}} & \multicolumn{1}{c}{1.16} & \multicolumn{1}{c}{0.99} & \multicolumn{1}{c}{1.02} & \multicolumn{1}{c}{\textbf{1.02}} & \multicolumn{1}{c}{\textbf{1.09}}  & \multicolumn{1}{c}{\textbf{1.1}}  & \multicolumn{1}{c}{\textbf{1.82}} & \multicolumn{1}{c}{\textbf{1.02}} & \multicolumn{1}{c}{\textbf{1.07}} & \multicolumn{1}{c}{\textbf{1.15}} & \multicolumn{1}{c}{\textbf{0.99}} & \multicolumn{1}{c}{\textbf{1}}    & \multicolumn{1}{c|}{\textbf{1.02}} &  &  &  \\ \cline{2-2}
\multicolumn{1}{|c|}{}                                        & \multicolumn{1}{c|}{Area}  & \multicolumn{1}{c}{1}    & \multicolumn{1}{c}{1}    & \multicolumn{1}{c}{1}    & \multicolumn{1}{c}{0.99} & \multicolumn{1}{c}{0.99} & \multicolumn{1}{c}{1}    & \multicolumn{1}{c}{0.99}  & \multicolumn{1}{c}{1}    & \multicolumn{1}{c}{1.03} & \multicolumn{1}{c}{1.04} & \multicolumn{1}{c}{1.08} & \multicolumn{1}{c}{1.14} & \multicolumn{1}{c}{1}    & \multicolumn{1}{c}{1}    & \multicolumn{1}{c|}{1}    &  &  &  \\ \cline{1-2}
\multicolumn{1}{|c|}{\multirow{3}{*}{TRANSPOSE (Reset only)}}     & \multicolumn{1}{c|}{PDP}   & \multicolumn{1}{c}{\textbf{0.98}} & \multicolumn{1}{c}{\textbf{0.97}} & \multicolumn{1}{c}{\textbf{0.97}} & \multicolumn{1}{c}{\textbf{0.92}} & \multicolumn{1}{c}{0.99} & \multicolumn{1}{c}{\textbf{1.03}} & \multicolumn{1}{c}{1.17}  & \multicolumn{1}{c}{1.34} & \multicolumn{1}{c}{1.36} & \multicolumn{1}{c}{\textbf{1.02}} & \multicolumn{1}{c}{1.28} & \multicolumn{1}{c}{1.57} & \multicolumn{1}{c}{0.99} & \multicolumn{1}{c}{1.02} & \multicolumn{1}{c|}{1.04} &  &  &  \\ \cline{2-2}
\multicolumn{1}{|c|}{}                                        & \multicolumn{1}{c|}{Power} & \multicolumn{1}{c}{1.13} & \multicolumn{1}{c}{1.14} & \multicolumn{1}{c}{1.15} & \multicolumn{1}{c}{\textbf{0.98}} & \multicolumn{1}{c}{1.01} & \multicolumn{1}{c}{1.05} & \multicolumn{1}{c}{1.7}   & \multicolumn{1}{c}{2.28} & \multicolumn{1}{c}{2.28} & \multicolumn{1}{c}{1.04} & \multicolumn{1}{c}{1.11} & \multicolumn{1}{c}{1.33} & \multicolumn{1}{c}{1}    & \multicolumn{1}{c}{1.02} & \multicolumn{1}{c|}{1.03} &  &  &  \\ \cline{2-2}
\multicolumn{1}{|c|}{}                                        & \multicolumn{1}{c|}{Area}  & \multicolumn{1}{c}{1}    & \multicolumn{1}{c}{1}    & \multicolumn{1}{c}{1}    & \multicolumn{1}{c}{0.99} & \multicolumn{1}{c}{1}    & \multicolumn{1}{c}{1}    & \multicolumn{1}{c}{1.04}  & \multicolumn{1}{c}{1.14} & \multicolumn{1}{c}{1.15} & \multicolumn{1}{c}{1.02} & \multicolumn{1}{c}{1.1}  & \multicolumn{1}{c}{1.18} & \multicolumn{1}{c}{1}    & \multicolumn{1}{c}{1.01} & \multicolumn{1}{c|}{1.01} &  &  &  \\ \cline{1-2}
\multicolumn{1}{|c|}{\multirow{3}{*}{TRANSPOSE (Set only)}}       & \multicolumn{1}{c|}{PDP}   & \multicolumn{1}{c}{0.99} & \multicolumn{1}{c}{0.99} & \multicolumn{1}{c}{0.99} & \multicolumn{1}{c}{1.02} & \multicolumn{1}{c}{1.09} & \multicolumn{1}{c}{1.1}  & \multicolumn{1}{c}{1.18}  & \multicolumn{1}{c}{1.33} & \multicolumn{1}{c}{1.35} & \multicolumn{1}{c}{1.08} & \multicolumn{1}{c}{1.25} & \multicolumn{1}{c}{1.55} & \multicolumn{1}{c}{1.01} & \multicolumn{1}{c}{1.02} & \multicolumn{1}{c|}{1.03} &  &  &  \\ \cline{2-2}
\multicolumn{1}{|c|}{}                                        & \multicolumn{1}{c|}{Power} & \multicolumn{1}{c}{1.1}  & \multicolumn{1}{c}{\textbf{1.12}} & \multicolumn{1}{c}{1.17} & \multicolumn{1}{c}{1}    & \multicolumn{1}{c}{1.04} & \multicolumn{1}{c}{1.04} & \multicolumn{1}{c}{1.65}  & \multicolumn{1}{c}{2.33} & \multicolumn{1}{c}{2.42} & \multicolumn{1}{c}{1.05} & \multicolumn{1}{c}{1.2}  & \multicolumn{1}{c}{1.35} & \multicolumn{1}{c}{1}    & \multicolumn{1}{c}{1.02} & \multicolumn{1}{c|}{\textbf{1.02}} &  &  &  \\ \cline{2-2}
\multicolumn{1}{|c|}{}                                        & \multicolumn{1}{c|}{Area}  & \multicolumn{1}{c}{1}    & \multicolumn{1}{c}{1}    & \multicolumn{1}{c}{1}    & \multicolumn{1}{c}{0.99} & \multicolumn{1}{c}{1}    & \multicolumn{1}{c}{1}    & \multicolumn{1}{c}{1.06}  & \multicolumn{1}{c}{1.12} & \multicolumn{1}{c}{1.13} & \multicolumn{1}{c}{1.06} & \multicolumn{1}{c}{1.14} & \multicolumn{1}{c}{1.18} & \multicolumn{1}{c}{0.99} & \multicolumn{1}{c}{1}    & \multicolumn{1}{c|}{1.02} &  &  &  \\ \cline{1-2}
\multicolumn{1}{|c|}{\multirow{3}{*}{TRANSPOSE (Set and Reset)}} & \multicolumn{1}{c|}{PDP}   & \multicolumn{1}{c}{0.99} & \multicolumn{1}{c}{1.05} & \multicolumn{1}{c}{1.05} & \multicolumn{1}{c}{\textbf{0.92}} & \multicolumn{1}{c}{\textbf{0.97}} & \multicolumn{1}{c}{\textbf{1.03}} & \multicolumn{1}{c}{1.17}  & \multicolumn{1}{c}{1.34} & \multicolumn{1}{c}{1.34} & \multicolumn{1}{c}{1.1}  & \multicolumn{1}{c}{1.29} & \multicolumn{1}{c}{1.67} & \multicolumn{1}{c}{1.03} & \multicolumn{1}{c}{1.04} & \multicolumn{1}{c|}{1.07} &  &  &  \\ \cline{2-2}
\multicolumn{1}{|c|}{}                                        & \multicolumn{1}{c|}{Power} & \multicolumn{1}{c}{1.12} & \multicolumn{1}{c}{\textbf{1.12}} & \multicolumn{1}{c}{\textbf{1.13}} & \multicolumn{1}{c}{0.99} & \multicolumn{1}{c}{\textbf{1}}    & \multicolumn{1}{c}{1.04} & \multicolumn{1}{c}{1.63}  & \multicolumn{1}{c}{1.91} & \multicolumn{1}{c}{2.41} & \multicolumn{1}{c}{1.15} & \multicolumn{1}{c}{1.21} & \multicolumn{1}{c}{1.44} & \multicolumn{1}{c}{1}    & \multicolumn{1}{c}{1.02} & \multicolumn{1}{c|}{1.03} &  &  &  \\ \cline{2-2}
\multicolumn{1}{|c|}{}                                        & \multicolumn{1}{c|}{Area}  & \multicolumn{1}{c}{1}    & \multicolumn{1}{c}{1}    & \multicolumn{1}{c}{1}    & \multicolumn{1}{c}{0.99} & \multicolumn{1}{c}{1}    & \multicolumn{1}{c}{1.02} & \multicolumn{1}{c}{1.07}  & \multicolumn{1}{c}{1.13} & \multicolumn{1}{c}{1.14} & \multicolumn{1}{c}{1.03} & \multicolumn{1}{c}{1.16} & \multicolumn{1}{c}{1.21} & \multicolumn{1}{c}{1.02} & \multicolumn{1}{c}{1.03} & \multicolumn{1}{c|}{1.03} &  &  &  \\ \cline{1-2}
\multicolumn{1}{|c|}{\multirow{3}{*}{PATRON (Average)}}       & \multicolumn{1}{c|}{PDP}   & \multicolumn{1}{c}{0.99} & \multicolumn{1}{c}{1.02} & \multicolumn{1}{c}{1.07} & \multicolumn{1}{c}{1}    & \multicolumn{1}{c}{\textbf{0.97}} & \multicolumn{1}{c}{1.04} & \multicolumn{1}{c}{1.17}  & \multicolumn{1}{c}{1.38} & \multicolumn{1}{c}{1.4}  & \multicolumn{1}{c}{1.19} & \multicolumn{1}{c}{1.35} & \multicolumn{1}{c}{1.75} & \multicolumn{1}{c}{1.02} & \multicolumn{1}{c}{1.04} & \multicolumn{1}{c|}{1.08} &  &  &  \\ \cline{2-2}
\multicolumn{1}{|c|}{}                                        & \multicolumn{1}{c|}{Power} & \multicolumn{1}{c}{1.13} & \multicolumn{1}{c}{1.14} & \multicolumn{1}{c}{1.18} & \multicolumn{1}{c}{0.99} & \multicolumn{1}{c}{1.04} & \multicolumn{1}{c}{1.05} & \multicolumn{1}{c}{1.67}  & \multicolumn{1}{c}{2.21} & \multicolumn{1}{c}{2.41} & \multicolumn{1}{c}{1.17} & \multicolumn{1}{c}{1.28} & \multicolumn{1}{c}{1.49} & \multicolumn{1}{c}{1}    & \multicolumn{1}{c}{1.03} & \multicolumn{1}{c|}{1.03} &  &  &  \\ \cline{2-2}
\multicolumn{1}{|c|}{}                                        & \multicolumn{1}{c|}{Area}  & \multicolumn{1}{c}{1}    & \multicolumn{1}{c}{1}    & \multicolumn{1}{c}{1}    & \multicolumn{1}{c}{0.99} & \multicolumn{1}{c}{1}    & \multicolumn{1}{c}{1.01} & \multicolumn{1}{c}{1.07}  & \multicolumn{1}{c}{1.14} & \multicolumn{1}{c}{1.17} & \multicolumn{1}{c}{1.1}  & \multicolumn{1}{c}{1.21} & \multicolumn{1}{c}{1.27} & \multicolumn{1}{c}{1.02} & \multicolumn{1}{c}{1.03} & \multicolumn{1}{c|}{1.04} &  &  &  \\ \cline{1-2}
\multicolumn{1}{|c|}{\multirow{3}{*}{SPARSE (Average)}}       & \multicolumn{1}{c|}{PDP}   & \multicolumn{1}{c}{1.01} & \multicolumn{1}{c}{1.01} & \multicolumn{1}{c}{1.1}  & \multicolumn{1}{c}{1.02} & \multicolumn{1}{c}{1.04} & \multicolumn{1}{c}{1.06} & \multicolumn{1}{c}{1.18} & \multicolumn{1}{c}{1.41} & \multicolumn{1}{c}{1.42} & \multicolumn{1}{c}{1.18} & \multicolumn{1}{c}{1.37} & \multicolumn{1}{c}{1.76} & \multicolumn{1}{c}{1.02} & \multicolumn{1}{c}{1.03} & \multicolumn{1}{c|}{1.09} &  &  &  \\ \cline{2-2}
\multicolumn{1}{|c|}{}                                        & \multicolumn{1}{c|}{Power} & \multicolumn{1}{c}{1.13} & \multicolumn{1}{c}{1.14} & \multicolumn{1}{c}{1.19} & \multicolumn{1}{c}{1.01} & \multicolumn{1}{c}{1.05} & \multicolumn{1}{c}{1.05} & \multicolumn{1}{c}{1.56}  & \multicolumn{1}{c}{2.19} & \multicolumn{1}{c}{2.52} & \multicolumn{1}{c}{1.19} & \multicolumn{1}{c}{1.28} & \multicolumn{1}{c}{1.52} & \multicolumn{1}{c}{1.03} & \multicolumn{1}{c}{1.03} & \multicolumn{1}{c|}{1.05} &  &  &  \\ \cline{2-2}
\multicolumn{1}{|c|}{}                                        & \multicolumn{1}{c|}{Area}  & \multicolumn{1}{c}{1}    & \multicolumn{1}{c}{1}    & \multicolumn{1}{c}{1}    & \multicolumn{1}{c}{1}    & \multicolumn{1}{c}{1.01} & \multicolumn{1}{c}{1.01} & \multicolumn{1}{c}{1.06}  & \multicolumn{1}{c}{1.12} & \multicolumn{1}{c}{1.2}  & \multicolumn{1}{c}{1.1}  & \multicolumn{1}{c}{1.18} & \multicolumn{1}{c}{1.29} & \multicolumn{1}{c}{1.04} & \multicolumn{1}{c}{1.05} & \multicolumn{1}{c|}{1.06} &  &  &  \\ \cline{1-2}
\multicolumn{1}{|c|}{\multirow{3}{*}{Codetables (Average)}}   & \multicolumn{1}{c|}{PDP}   & \multicolumn{1}{c}{0.99} & \multicolumn{1}{c}{1.03} & \multicolumn{1}{c}{1.07} & \multicolumn{1}{c}{0.99} & \multicolumn{1}{c}{1.1}  & \multicolumn{1}{c}{1.1}  & \multicolumn{1}{c}{1.18}  & \multicolumn{1}{c}{1.36} & \multicolumn{1}{c}{1.39} & \multicolumn{1}{c}{1.18} & \multicolumn{1}{c}{1.38} & \multicolumn{1}{c}{1.78} & \multicolumn{1}{c}{1.03} & \multicolumn{1}{c}{1.05} & \multicolumn{1}{c|}{1.08} &  &  &  \\ \cline{2-2}
\multicolumn{1}{|c|}{}                                        & \multicolumn{1}{c|}{Power} & \multicolumn{1}{c}{1.13} & \multicolumn{1}{c}{1.14} & \multicolumn{1}{c}{1.17} & \multicolumn{1}{c}{1.01} & \multicolumn{1}{c}{1.05} & \multicolumn{1}{c}{1.06} & \multicolumn{1}{c}{1.65}  & \multicolumn{1}{c}{2.29} & \multicolumn{1}{c}{2.43} & \multicolumn{1}{c}{1.16} & \multicolumn{1}{c}{1.26} & \multicolumn{1}{c}{1.51} & \multicolumn{1}{c}{1.01} & \multicolumn{1}{c}{1.04} & \multicolumn{1}{c|}{1.04} &  &  &  \\ \cline{2-2}
\multicolumn{1}{|c|}{}                                        & \multicolumn{1}{c|}{Area}  & \multicolumn{1}{c}{1}    & \multicolumn{1}{c}{1}    & \multicolumn{1}{c}{1}    & \multicolumn{1}{c}{0.99} & \multicolumn{1}{c}{1}    & \multicolumn{1}{c}{1.03} & \multicolumn{1}{c}{1.06}  & \multicolumn{1}{c}{1.13} & \multicolumn{1}{c}{1.15} & \multicolumn{1}{c}{1.07} & \multicolumn{1}{c}{1.15} & \multicolumn{1}{c}{1.28} & \multicolumn{1}{c}{1.03} & \multicolumn{1}{c}{1.04} & \multicolumn{1}{c|}{1.04} &  &  &  \\ \cline{1-17}
\end{tabular}%
}
\caption{Power-delay product (PDP), FSM module power, and area analysis for various encoding schemes. All PDP and area values are \textbf{normalized} with respect to Binary Encoding. The averages of five values are considered for PATRON, SPARSE, and Codetables. The minimum PDP and power for each $x$ are in bold.}
\label{tableresults}
\end{table*}

\begin{figure*}[] 
\centering
\subfloat[]{\includegraphics[clip,width=0.33\textwidth]{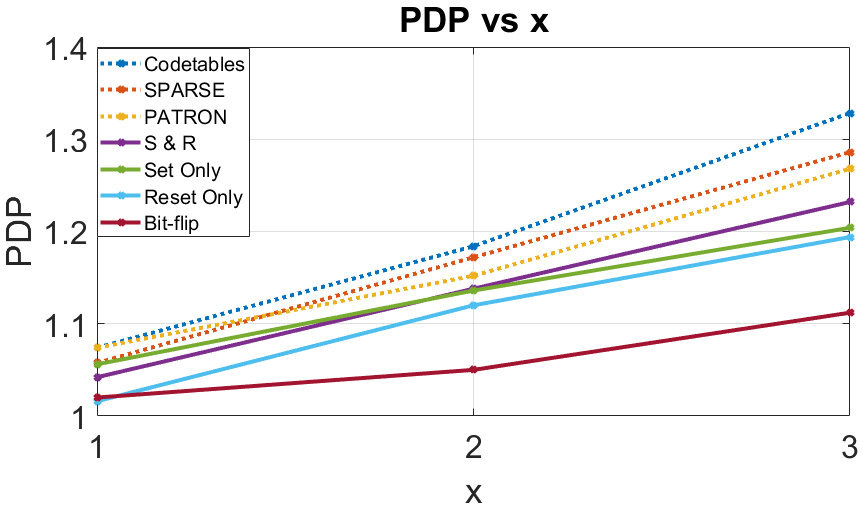}} \hspace{0.05em}%
\subfloat[]{\includegraphics[clip,width=0.33\textwidth]{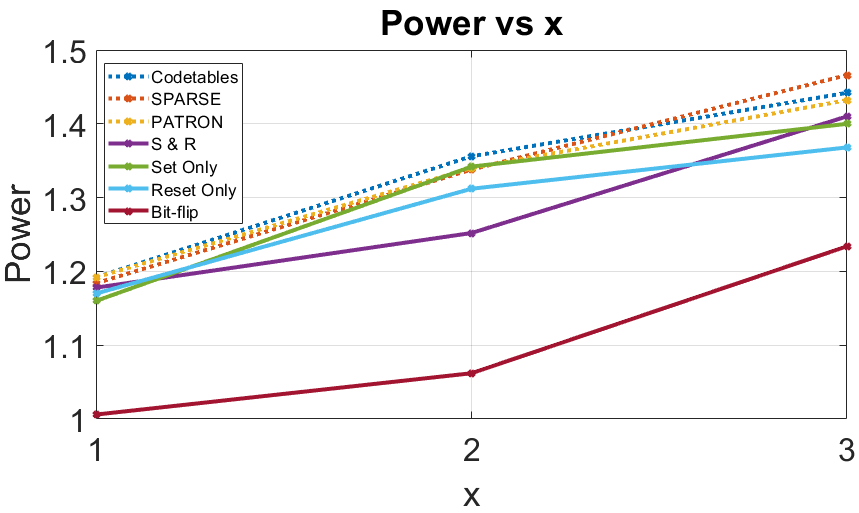}} 
\hspace{0.05em}%
\subfloat[]{\includegraphics[clip,width=0.33\textwidth]{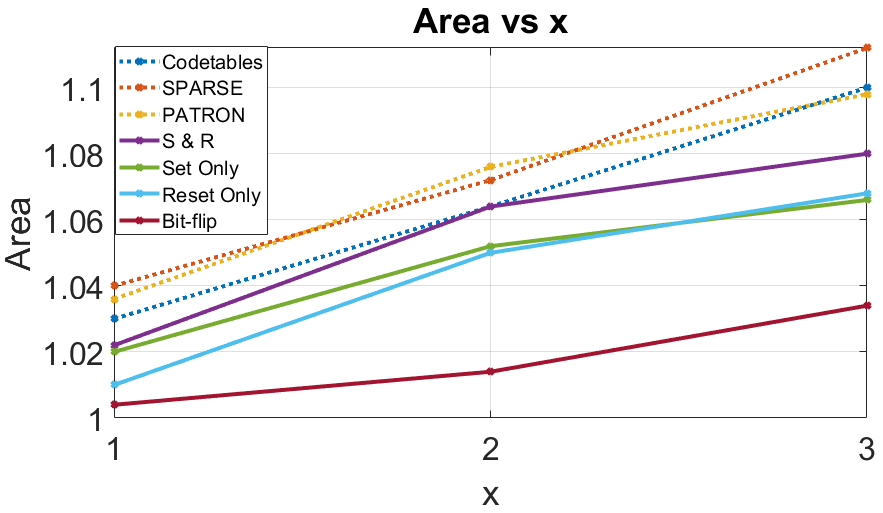}}
\caption{Normalized PDP, Power, and Area for different approaches averaged for all benchmarks.}
\label{comparingpdpareapower}
\end{figure*}

 \vspace{0.5ex}
\noindent \textbf{PDP, Power and Area Overhead Comparison:}
Table \ref{tableresults}  provides a detailed comparative analysis of TRANSPOSE against other  \emph{model-unaware} approaches. Pertinent details for each benchmark are summarized in Table  \ref{firsttable}.  Additionally,  Fig. \ref{comparingpdpareapower} visually depicts the variation of these metrics across different approaches.
Since the PDP, power, and area values are normalized against binary encoding, PATRON, SPARSE, and Codetables produce encodings that are not optimized for power efficiency due to the absence of criteria for selecting the optimal encoding that meets specific FSM design specifications.

As anticipated, Codetables' linear encoding yields the highest average PDPs, primarily because the solution is strictly guided by the Hamming Distance (HD) without considering other crucial FSM design specifications such as transitional probabilities and relaxed constraints of ($\mathbb{T}-\mathbb{AT}$) transitions. SPARSE and PATRON exhibit the next highest PDPs, respectively. Variations in PDP among these approaches can be attributed to how they handle designer-defined sensitive states and the protection of \emph{all} $SS$ from $NS$ choices, including each state within SS, leading to redundant protection mechanisms in the design. SPARSE slightly outperforms PATRON in PDP, likely due to its handling of initialized sensitive states. TRANSPOSE approaches demonstrate superior PDPs because they offer the flexibility to generate \emph{a single encoding solution}  that precisely aligns with FSM design parameters based on different models. Equally significant is their power optimization capability and the relaxed handling of  ($\mathbb{T}-\mathbb{AT}$) transitions.  Specifically, TRANSPOSE (bit flip model) and the reset-only approach within the set-reset model achieve better PDPs compared to model-unaware approaches. An interesting comparison arises between TRANSPOSE (Bit-Flip) and SPARSE, both utilizing the bit-flip model~\cite{choudhury2021sparse}. TRANSPOSE (Bit-Flip) outperforms SPARSE due to its power optimization step and the flexibility to selectively protect certain $\mathbb{AT}$ transitions, which optimizes the number of flip-flops ($n$). 

In terms of individual power and area, TRANSPOSE exhibits the lowest overall average overhead compared to other model-unaware approaches. This outcome is attributed to its optimization flow, which strives to minimize the required number of FFs. Note that, although for TRANSPOSE (Bit-Flip) approach, all the $\mathbb{FF}$ corresponding to only the FSM encoded module are placed a secure distance apart (not the sensitive FF regions), the normalized design area still remains minimum as shown in Fig.~\ref{comparingpdpareapower}(c). The benchmarks are intentionally chosen to encompass a broad spectrum of ratios between the size of the FSM-encoded module and the complete design. 
On average, the individual power consumption of the FSM encoded module compared to the entire design is observed to be $22.7\%$ (min: AES (0.3\%), max: Power Sequencer (59\%)). From the power-area correlation, constraining area by placing the sensitive regions of the $\mathbb{SFF}$ for all these variably-sized FSMs in relation to the whole design still places TRANSPOSE approaches ahead of the state-based approaches. 
On average, compared to the state-based approaches the TRANSPOSE approaches are seen to be less by 5.5\% in PDP, 6.46\% in power, and  2.75\% in area.

  \begin{table*}[]
\centering
\scalebox{0.92}
{\begin{tabular}{llllllllllllllllllll}
\cline{3-17}
\multicolumn{1}{c}{}                                         & \multicolumn{1}{c|}{}             & \multicolumn{3}{c|}{AES}                                                                                                                                                                                                 & \multicolumn{3}{c|}{SHA-256}                                                                                                                                                                                             & \multicolumn{3}{c|}{FSM   Controller}                                                                                                                                                                                    & \multicolumn{3}{c|}{Power   Sequencer}                                                                                                                                                                                   & \multicolumn{3}{c|}{VIIRF}                                                                                                                                                                                                  &  &  &  \\ \cline{2-17}
\multicolumn{1}{c|}{}                                        & \multicolumn{1}{c|}{x}            & \multicolumn{1}{c|}{1}                                                 & \multicolumn{1}{c|}{2}                                                 & \multicolumn{1}{c|}{3}                                                 & \multicolumn{1}{c|}{1}                                                 & \multicolumn{1}{c|}{2}                                                 & \multicolumn{1}{c|}{3}                                                 & \multicolumn{1}{c|}{1}                                                 & \multicolumn{1}{c|}{2}                                                 & \multicolumn{1}{c|}{3}                                                 & \multicolumn{1}{c|}{1}                                                 & \multicolumn{1}{c|}{2}                                                 & \multicolumn{1}{c|}{3}                                                 & \multicolumn{1}{c|}{1}                                                  & \multicolumn{1}{c|}{2}                                                  & \multicolumn{1}{c|}{3}                                                  &  &  &  \\ \cline{1-17}
\multicolumn{1}{|c|}{}                                       & \multicolumn{1}{c|}{$VM$}           & \multicolumn{1}{c}{\cellcolor[HTML]{FFC7CE}{\color[HTML]{9C0006} 0.2}} & \multicolumn{1}{c}{\cellcolor[HTML]{FFC7CE}{\color[HTML]{9C0006} 0.4}} & \multicolumn{1}{c}{\cellcolor[HTML]{FFC7CE}{\color[HTML]{9C0006} 0.4}} & \multicolumn{1}{c}{\cellcolor[HTML]{FFC7CE}{\color[HTML]{9C0006} 0.6}} & \multicolumn{1}{c}{\cellcolor[HTML]{FFC7CE}{\color[HTML]{9C0006} 0.6}} & \multicolumn{1}{c}{\cellcolor[HTML]{FFC7CE}{\color[HTML]{9C0006} 0.6}} & \multicolumn{1}{c}{\cellcolor[HTML]{FFC7CE}{\color[HTML]{9C0006} 0.3}} & \multicolumn{1}{c}{\cellcolor[HTML]{FFC7CE}{\color[HTML]{9C0006} 0.5}} & \multicolumn{1}{c}{\cellcolor[HTML]{FFC7CE}{\color[HTML]{9C0006} 0.5}} & \multicolumn{1}{c}{\cellcolor[HTML]{FFC7CE}{\color[HTML]{9C0006} 0.3}} & \multicolumn{1}{c}{\cellcolor[HTML]{FFC7CE}{\color[HTML]{9C0006} 0.3}} & \multicolumn{1}{c}{\cellcolor[HTML]{FFC7CE}{\color[HTML]{9C0006} 0.4}} & \multicolumn{1}{c}{\cellcolor[HTML]{FFC7CE}{\color[HTML]{9C0006} 0.2}}  & \multicolumn{1}{c}{\cellcolor[HTML]{FFC7CE}{\color[HTML]{9C0006} 0.3}}  & \multicolumn{1}{c|}{\cellcolor[HTML]{FFC7CE}{\color[HTML]{9C0006} 0.4}}  &  &  &  \\ \cline{2-2}
\multicolumn{1}{|c|}{}                                       & \multicolumn{1}{c|}{$SVM$} & \multicolumn{1}{c}{\cellcolor[HTML]{C6EFCE}{\color[HTML]{006100} 0}}   & \multicolumn{1}{c}{\cellcolor[HTML]{C6EFCE}{\color[HTML]{006100} 0}}   & \multicolumn{1}{c}{\cellcolor[HTML]{C6EFCE}{\color[HTML]{006100} 0}}   & \multicolumn{1}{c}{\cellcolor[HTML]{C6EFCE}{\color[HTML]{006100} 0}}   & \multicolumn{1}{c}{\cellcolor[HTML]{C6EFCE}{\color[HTML]{006100} 0}}   & \multicolumn{1}{c}{\cellcolor[HTML]{C6EFCE}{\color[HTML]{006100} 0}}   & \multicolumn{1}{c}{\cellcolor[HTML]{C6EFCE}{\color[HTML]{006100} 0}}   & \multicolumn{1}{c}{\cellcolor[HTML]{C6EFCE}{\color[HTML]{006100} 0}}   & \multicolumn{1}{c}{\cellcolor[HTML]{C6EFCE}{\color[HTML]{006100} 0}}   & \multicolumn{1}{c}{\cellcolor[HTML]{C6EFCE}{\color[HTML]{006100} 0}}   & \multicolumn{1}{c}{\cellcolor[HTML]{C6EFCE}{\color[HTML]{006100} 0}}   & \multicolumn{1}{c}{\cellcolor[HTML]{C6EFCE}{\color[HTML]{006100} 0}}   & \multicolumn{1}{c}{\cellcolor[HTML]{C6EFCE}{\color[HTML]{006100} 0}}    & \multicolumn{1}{c}{\cellcolor[HTML]{C6EFCE}{\color[HTML]{006100} 0}}    & \multicolumn{1}{c|}{\cellcolor[HTML]{C6EFCE}{\color[HTML]{006100} 0}}    &  &  &  \\ \cline{2-2}
\multicolumn{1}{|c|}{\multirow{-3}{*}{TRANSPOSE (Bit-Flip)}}     & \multicolumn{1}{c|}{$STVM_{bf}$}       & \multicolumn{1}{c}{\cellcolor[HTML]{C6EFCE}{\color[HTML]{006100} 0}}   & \multicolumn{1}{c}{\cellcolor[HTML]{C6EFCE}{\color[HTML]{006100} 0}}   & \multicolumn{1}{c}{\cellcolor[HTML]{C6EFCE}{\color[HTML]{006100} 0}}   & \multicolumn{1}{c}{\cellcolor[HTML]{C6EFCE}{\color[HTML]{006100} 0}}   & \multicolumn{1}{c}{\cellcolor[HTML]{C6EFCE}{\color[HTML]{006100} 0}}   & \multicolumn{1}{c}{\cellcolor[HTML]{C6EFCE}{\color[HTML]{006100} 0}}   & \multicolumn{1}{c}{\cellcolor[HTML]{C6EFCE}{\color[HTML]{006100} 0}}   & \multicolumn{1}{c}{\cellcolor[HTML]{C6EFCE}{\color[HTML]{006100} 0}}   & \multicolumn{1}{c}{\cellcolor[HTML]{C6EFCE}{\color[HTML]{006100} 0}}   & \multicolumn{1}{c}{\cellcolor[HTML]{C6EFCE}{\color[HTML]{006100} 0}}   & \multicolumn{1}{c}{\cellcolor[HTML]{C6EFCE}{\color[HTML]{006100} 0}}   & \multicolumn{1}{c}{\cellcolor[HTML]{C6EFCE}{\color[HTML]{006100} 0}}   & \multicolumn{1}{c}{\cellcolor[HTML]{C6EFCE}{\color[HTML]{006100} 0}}    & \multicolumn{1}{c}{\cellcolor[HTML]{C6EFCE}{\color[HTML]{006100} 0}}    & \multicolumn{1}{c|}{\cellcolor[HTML]{C6EFCE}{\color[HTML]{006100} 0}}    &  &  &  \\ \cline{1-2}
\multicolumn{1}{|c|}{}                                       & \multicolumn{1}{c|}{$VM$}           & \multicolumn{1}{c}{\cellcolor[HTML]{FFC7CE}{\color[HTML]{9C0006} 0.2}} & \multicolumn{1}{c}{\cellcolor[HTML]{FFC7CE}{\color[HTML]{9C0006} 0.4}} & \multicolumn{1}{c}{\cellcolor[HTML]{FFC7CE}{\color[HTML]{9C0006} 0.4}} & \multicolumn{1}{c}{\cellcolor[HTML]{FFC7CE}{\color[HTML]{9C0006} 0.3}} & \multicolumn{1}{c}{\cellcolor[HTML]{FFC7CE}{\color[HTML]{9C0006} 0.4}} & \multicolumn{1}{c}{\cellcolor[HTML]{FFC7CE}{\color[HTML]{9C0006} 0.4}} & \multicolumn{1}{c}{\cellcolor[HTML]{FFC7CE}{\color[HTML]{9C0006} 0.3}} & \multicolumn{1}{c}{\cellcolor[HTML]{FFC7CE}{\color[HTML]{9C0006} 0.5}} & \multicolumn{1}{c}{\cellcolor[HTML]{FFC7CE}{\color[HTML]{9C0006} 0.5}} & \multicolumn{1}{c}{\cellcolor[HTML]{FFC7CE}{\color[HTML]{9C0006} 0.3}} & \multicolumn{1}{c}{\cellcolor[HTML]{FFC7CE}{\color[HTML]{9C0006} 0.3}} & \multicolumn{1}{c}{\cellcolor[HTML]{FFC7CE}{\color[HTML]{9C0006} 0.4}} & \multicolumn{1}{c}{\cellcolor[HTML]{FFC7CE}{\color[HTML]{9C0006} 0.2}}  & \multicolumn{1}{c}{\cellcolor[HTML]{FFC7CE}{\color[HTML]{9C0006} 0.3}}  & \multicolumn{1}{c|}{\cellcolor[HTML]{FFC7CE}{\color[HTML]{9C0006} 0.3}}  &  &  &  \\ \cline{2-2}
\multicolumn{1}{|c|}{}                                       & \multicolumn{1}{c|}{$SVM$} & \multicolumn{1}{c}{\cellcolor[HTML]{C6EFCE}{\color[HTML]{006100} 0}}   & \multicolumn{1}{c}{\cellcolor[HTML]{C6EFCE}{\color[HTML]{006100} 0}}   & \multicolumn{1}{c}{\cellcolor[HTML]{C6EFCE}{\color[HTML]{006100} 0}}   & \multicolumn{1}{c}{\cellcolor[HTML]{C6EFCE}{\color[HTML]{006100} 0}}   & \multicolumn{1}{c}{\cellcolor[HTML]{C6EFCE}{\color[HTML]{006100} 0}}   & \multicolumn{1}{c}{\cellcolor[HTML]{C6EFCE}{\color[HTML]{006100} 0}}   & \multicolumn{1}{c}{\cellcolor[HTML]{C6EFCE}{\color[HTML]{006100} 0}}   & \multicolumn{1}{c}{\cellcolor[HTML]{C6EFCE}{\color[HTML]{006100} 0}}   & \multicolumn{1}{c}{\cellcolor[HTML]{C6EFCE}{\color[HTML]{006100} 0}}   & \multicolumn{1}{c}{\cellcolor[HTML]{C6EFCE}{\color[HTML]{006100} 0}}   & \multicolumn{1}{c}{\cellcolor[HTML]{C6EFCE}{\color[HTML]{006100} 0}}   & \multicolumn{1}{c}{\cellcolor[HTML]{C6EFCE}{\color[HTML]{006100} 0}}   & \multicolumn{1}{c}{\cellcolor[HTML]{C6EFCE}{\color[HTML]{006100} 0}}    & \multicolumn{1}{c}{\cellcolor[HTML]{C6EFCE}{\color[HTML]{006100} 0}}    & \multicolumn{1}{c|}{\cellcolor[HTML]{C6EFCE}{\color[HTML]{006100} 0}}    &  &  &  \\ \cline{2-2}
\multicolumn{1}{|c|}{\multirow{-3}{*}{TRANSPOSE (Reset only)}}    & \multicolumn{1}{c|}{$STVM_{sr}$}       & \multicolumn{1}{c}{\cellcolor[HTML]{C6EFCE}{\color[HTML]{006100} 0}}   & \multicolumn{1}{c}{\cellcolor[HTML]{C6EFCE}{\color[HTML]{006100} 0}}   & \multicolumn{1}{c}{\cellcolor[HTML]{C6EFCE}{\color[HTML]{006100} 0}}   & \multicolumn{1}{c}{\cellcolor[HTML]{C6EFCE}{\color[HTML]{006100} 0}}   & \multicolumn{1}{c}{\cellcolor[HTML]{C6EFCE}{\color[HTML]{006100} 0}}   & \multicolumn{1}{c}{\cellcolor[HTML]{C6EFCE}{\color[HTML]{006100} 0}}   & \multicolumn{1}{c}{\cellcolor[HTML]{C6EFCE}{\color[HTML]{006100} 0}}   & \multicolumn{1}{c}{\cellcolor[HTML]{C6EFCE}{\color[HTML]{006100} 0}}   & \multicolumn{1}{c}{\cellcolor[HTML]{C6EFCE}{\color[HTML]{006100} 0}}   & \multicolumn{1}{c}{\cellcolor[HTML]{C6EFCE}{\color[HTML]{006100} 0}}   & \multicolumn{1}{c}{\cellcolor[HTML]{C6EFCE}{\color[HTML]{006100} 0}}   & \multicolumn{1}{c}{\cellcolor[HTML]{C6EFCE}{\color[HTML]{006100} 0}}   & \multicolumn{1}{c}{\cellcolor[HTML]{C6EFCE}{\color[HTML]{006100} 0}}    & \multicolumn{1}{c}{\cellcolor[HTML]{C6EFCE}{\color[HTML]{006100} 0}}    & \multicolumn{1}{c|}{\cellcolor[HTML]{C6EFCE}{\color[HTML]{006100} 0}}    &  &  &  \\ \cline{1-2}
\multicolumn{1}{|c|}{}                                       & \multicolumn{1}{c|}{$VM$}           & \multicolumn{1}{c}{\cellcolor[HTML]{FFC7CE}{\color[HTML]{9C0006} 0.2}} & \multicolumn{1}{c}{\cellcolor[HTML]{FFC7CE}{\color[HTML]{9C0006} 0.4}} & \multicolumn{1}{c}{\cellcolor[HTML]{FFC7CE}{\color[HTML]{9C0006} 0.4}} & \multicolumn{1}{c}{\cellcolor[HTML]{FFC7CE}{\color[HTML]{9C0006} 0.1}} & \multicolumn{1}{c}{\cellcolor[HTML]{FFC7CE}{\color[HTML]{9C0006} 0.6}} & \multicolumn{1}{c}{\cellcolor[HTML]{FFC7CE}{\color[HTML]{9C0006} 0.6}} & \multicolumn{1}{c}{\cellcolor[HTML]{FFC7CE}{\color[HTML]{9C0006} 0.3}} & \multicolumn{1}{c}{\cellcolor[HTML]{FFC7CE}{\color[HTML]{9C0006} 0.5}} & \multicolumn{1}{c}{\cellcolor[HTML]{FFC7CE}{\color[HTML]{9C0006} 0.5}} & \multicolumn{1}{c}{\cellcolor[HTML]{FFC7CE}{\color[HTML]{9C0006} 0.3}} & \multicolumn{1}{c}{\cellcolor[HTML]{FFC7CE}{\color[HTML]{9C0006} 0.3}} & \multicolumn{1}{c}{\cellcolor[HTML]{FFC7CE}{\color[HTML]{9C0006} 0.4}} & \multicolumn{1}{c}{\cellcolor[HTML]{FFC7CE}{\color[HTML]{9C0006} 0.2}}  & \multicolumn{1}{c}{\cellcolor[HTML]{FFC7CE}{\color[HTML]{9C0006} 0.4}}  & \multicolumn{1}{c|}{\cellcolor[HTML]{FFC7CE}{\color[HTML]{9C0006} 0.4}}  &  &  &  \\ \cline{2-2}
\multicolumn{1}{|c|}{}                                       & \multicolumn{1}{c|}{$SVM$} & \multicolumn{1}{c}{\cellcolor[HTML]{C6EFCE}{\color[HTML]{006100} 0}}   & \multicolumn{1}{c}{\cellcolor[HTML]{C6EFCE}{\color[HTML]{006100} 0}}   & \multicolumn{1}{c}{\cellcolor[HTML]{C6EFCE}{\color[HTML]{006100} 0}}   & \multicolumn{1}{c}{\cellcolor[HTML]{C6EFCE}{\color[HTML]{006100} 0}}   & \multicolumn{1}{c}{\cellcolor[HTML]{C6EFCE}{\color[HTML]{006100} 0}}   & \multicolumn{1}{c}{\cellcolor[HTML]{C6EFCE}{\color[HTML]{006100} 0}}   & \multicolumn{1}{c}{\cellcolor[HTML]{C6EFCE}{\color[HTML]{006100} 0}}   & \multicolumn{1}{c}{\cellcolor[HTML]{C6EFCE}{\color[HTML]{006100} 0}}   & \multicolumn{1}{c}{\cellcolor[HTML]{C6EFCE}{\color[HTML]{006100} 0}}   & \multicolumn{1}{c}{\cellcolor[HTML]{C6EFCE}{\color[HTML]{006100} 0}}   & \multicolumn{1}{c}{\cellcolor[HTML]{C6EFCE}{\color[HTML]{006100} 0}}   & \multicolumn{1}{c}{\cellcolor[HTML]{C6EFCE}{\color[HTML]{006100} 0}}   & \multicolumn{1}{c}{\cellcolor[HTML]{C6EFCE}{\color[HTML]{006100} 0}}    & \multicolumn{1}{c}{\cellcolor[HTML]{C6EFCE}{\color[HTML]{006100} 0}}    & \multicolumn{1}{c|}{\cellcolor[HTML]{C6EFCE}{\color[HTML]{006100} 0}}    &  &  &  \\ \cline{2-2}
\multicolumn{1}{|c|}{\multirow{-3}{*}{TRANSPOSE (Set only)}}      & \multicolumn{1}{c|}{$STVM_{sr}$}       & \multicolumn{1}{c}{\cellcolor[HTML]{C6EFCE}{\color[HTML]{006100} 0}}   & \multicolumn{1}{c}{\cellcolor[HTML]{C6EFCE}{\color[HTML]{006100} 0}}   & \multicolumn{1}{c}{\cellcolor[HTML]{C6EFCE}{\color[HTML]{006100} 0}}   & \multicolumn{1}{c}{\cellcolor[HTML]{C6EFCE}{\color[HTML]{006100} 0}}   & \multicolumn{1}{c}{\cellcolor[HTML]{C6EFCE}{\color[HTML]{006100} 0}}   & \multicolumn{1}{c}{\cellcolor[HTML]{C6EFCE}{\color[HTML]{006100} 0}}   & \multicolumn{1}{c}{\cellcolor[HTML]{C6EFCE}{\color[HTML]{006100} 0}}   & \multicolumn{1}{c}{\cellcolor[HTML]{C6EFCE}{\color[HTML]{006100} 0}}   & \multicolumn{1}{c}{\cellcolor[HTML]{C6EFCE}{\color[HTML]{006100} 0}}   & \multicolumn{1}{c}{\cellcolor[HTML]{C6EFCE}{\color[HTML]{006100} 0}}   & \multicolumn{1}{c}{\cellcolor[HTML]{C6EFCE}{\color[HTML]{006100} 0}}   & \multicolumn{1}{c}{\cellcolor[HTML]{C6EFCE}{\color[HTML]{006100} 0}}   & \multicolumn{1}{c}{\cellcolor[HTML]{C6EFCE}{\color[HTML]{006100} 0}}    & \multicolumn{1}{c}{\cellcolor[HTML]{C6EFCE}{\color[HTML]{006100} 0}}    & \multicolumn{1}{c|}{\cellcolor[HTML]{C6EFCE}{\color[HTML]{006100} 0}}    &  &  &  \\ \cline{1-2}
\multicolumn{1}{|c|}{}                                       & \multicolumn{1}{c|}{$VM$}           & \multicolumn{1}{c}{\cellcolor[HTML]{FFC7CE}{\color[HTML]{9C0006} 0.2}} & \multicolumn{1}{c}{\cellcolor[HTML]{FFC7CE}{\color[HTML]{9C0006} 0.4}} & \multicolumn{1}{c}{\cellcolor[HTML]{FFC7CE}{\color[HTML]{9C0006} 0.4}} & \multicolumn{1}{c}{\cellcolor[HTML]{FFC7CE}{\color[HTML]{9C0006} 0.3}} & \multicolumn{1}{c}{\cellcolor[HTML]{FFC7CE}{\color[HTML]{9C0006} 0.4}} & \multicolumn{1}{c}{\cellcolor[HTML]{FFC7CE}{\color[HTML]{9C0006} 0.4}} & \multicolumn{1}{c}{\cellcolor[HTML]{FFC7CE}{\color[HTML]{9C0006} 0.4}} & \multicolumn{1}{c}{\cellcolor[HTML]{FFC7CE}{\color[HTML]{9C0006} 0.5}} & \multicolumn{1}{c}{\cellcolor[HTML]{FFC7CE}{\color[HTML]{9C0006} 0.5}} & \multicolumn{1}{c}{\cellcolor[HTML]{FFC7CE}{\color[HTML]{9C0006} 0.3}} & \multicolumn{1}{c}{\cellcolor[HTML]{FFC7CE}{\color[HTML]{9C0006} 0.3}} & \multicolumn{1}{c}{\cellcolor[HTML]{FFC7CE}{\color[HTML]{9C0006} 0.4}} & \multicolumn{1}{c}{\cellcolor[HTML]{FFC7CE}{\color[HTML]{9C0006} 0.2}}  & \multicolumn{1}{c}{\cellcolor[HTML]{FFC7CE}{\color[HTML]{9C0006} 0.4}}  & \multicolumn{1}{c|}{\cellcolor[HTML]{FFC7CE}{\color[HTML]{9C0006} 0.4}}  &  &  &  \\ \cline{2-2}
\multicolumn{1}{|c|}{}                                       & \multicolumn{1}{c|}{$SVM$} & \multicolumn{1}{c}{\cellcolor[HTML]{C6EFCE}{\color[HTML]{006100} 0}}   & \multicolumn{1}{c}{\cellcolor[HTML]{C6EFCE}{\color[HTML]{006100} 0}}   & \multicolumn{1}{c}{\cellcolor[HTML]{C6EFCE}{\color[HTML]{006100} 0}}   & \multicolumn{1}{c}{\cellcolor[HTML]{C6EFCE}{\color[HTML]{006100} 0}}   & \multicolumn{1}{c}{\cellcolor[HTML]{C6EFCE}{\color[HTML]{006100} 0}}   & \multicolumn{1}{c}{\cellcolor[HTML]{C6EFCE}{\color[HTML]{006100} 0}}   & \multicolumn{1}{c}{\cellcolor[HTML]{C6EFCE}{\color[HTML]{006100} 0}}   & \multicolumn{1}{c}{\cellcolor[HTML]{C6EFCE}{\color[HTML]{006100} 0}}   & \multicolumn{1}{c}{\cellcolor[HTML]{C6EFCE}{\color[HTML]{006100} 0}}   & \multicolumn{1}{c}{\cellcolor[HTML]{C6EFCE}{\color[HTML]{006100} 0}}   & \multicolumn{1}{c}{\cellcolor[HTML]{C6EFCE}{\color[HTML]{006100} 0}}   & \multicolumn{1}{c}{\cellcolor[HTML]{C6EFCE}{\color[HTML]{006100} 0}}   & \multicolumn{1}{c}{\cellcolor[HTML]{C6EFCE}{\color[HTML]{006100} 0}}    & \multicolumn{1}{c}{\cellcolor[HTML]{C6EFCE}{\color[HTML]{006100} 0}}    & \multicolumn{1}{c|}{\cellcolor[HTML]{C6EFCE}{\color[HTML]{006100} 0}}    &  &  &  \\ \cline{2-2}
\multicolumn{1}{|c|}{\multirow{-3}{*}{TRANSPOSE (Set and Reset)}} & \multicolumn{1}{c|}{$STVM_{sr}$}       & \multicolumn{1}{c}{\cellcolor[HTML]{C6EFCE}{\color[HTML]{006100} 0}}   & \multicolumn{1}{c}{\cellcolor[HTML]{C6EFCE}{\color[HTML]{006100} 0}}   & \multicolumn{1}{c}{\cellcolor[HTML]{C6EFCE}{\color[HTML]{006100} 0}}   & \multicolumn{1}{c}{\cellcolor[HTML]{C6EFCE}{\color[HTML]{006100} 0}}   & \multicolumn{1}{c}{\cellcolor[HTML]{C6EFCE}{\color[HTML]{006100} 0}}   & \multicolumn{1}{c}{\cellcolor[HTML]{C6EFCE}{\color[HTML]{006100} 0}}   & \multicolumn{1}{c}{\cellcolor[HTML]{C6EFCE}{\color[HTML]{006100} 0}}   & \multicolumn{1}{c}{\cellcolor[HTML]{C6EFCE}{\color[HTML]{006100} 0}}   & \multicolumn{1}{c}{\cellcolor[HTML]{C6EFCE}{\color[HTML]{006100} 0}}   & \multicolumn{1}{c}{\cellcolor[HTML]{C6EFCE}{\color[HTML]{006100} 0}}   & \multicolumn{1}{c}{\cellcolor[HTML]{C6EFCE}{\color[HTML]{006100} 0}}   & \multicolumn{1}{c}{\cellcolor[HTML]{C6EFCE}{\color[HTML]{006100} 0}}   & \multicolumn{1}{c}{\cellcolor[HTML]{C6EFCE}{\color[HTML]{006100} 0}}    & \multicolumn{1}{c}{\cellcolor[HTML]{C6EFCE}{\color[HTML]{006100} 0}}    & \multicolumn{1}{c|}{\cellcolor[HTML]{C6EFCE}{\color[HTML]{006100} 0}}    &  &  &  \\ \cline{1-2}
\multicolumn{1}{|c|}{}                                       & \multicolumn{1}{c|}{$VM$}           & \multicolumn{1}{c}{\cellcolor[HTML]{C6EFCE}{\color[HTML]{006100} 0}}   & \multicolumn{1}{c}{\cellcolor[HTML]{C6EFCE}{\color[HTML]{006100} 0}}   & \multicolumn{1}{c}{\cellcolor[HTML]{C6EFCE}{\color[HTML]{006100} 0}}   & \multicolumn{1}{c}{\cellcolor[HTML]{C6EFCE}{\color[HTML]{006100} 0}}   & \multicolumn{1}{c}{\cellcolor[HTML]{C6EFCE}{\color[HTML]{006100} 0}}   & \multicolumn{1}{c}{\cellcolor[HTML]{C6EFCE}{\color[HTML]{006100} 0}}   & \multicolumn{1}{c}{\cellcolor[HTML]{C6EFCE}{\color[HTML]{006100} 0}}   & \multicolumn{1}{c}{\cellcolor[HTML]{C6EFCE}{\color[HTML]{006100} 0}}   & \multicolumn{1}{c}{\cellcolor[HTML]{C6EFCE}{\color[HTML]{006100} 0}}   & \multicolumn{1}{c}{\cellcolor[HTML]{C6EFCE}{\color[HTML]{006100} 0}}   & \multicolumn{1}{c}{\cellcolor[HTML]{C6EFCE}{\color[HTML]{006100} 0}}   & \multicolumn{1}{c}{\cellcolor[HTML]{C6EFCE}{\color[HTML]{006100} 0}}   & \multicolumn{1}{c}{\cellcolor[HTML]{C6EFCE}{\color[HTML]{006100} 0}}    & \multicolumn{1}{c}{\cellcolor[HTML]{C6EFCE}{\color[HTML]{006100} 0}}    & \multicolumn{1}{c|}{\cellcolor[HTML]{C6EFCE}{\color[HTML]{006100} 0}}    &  &  &  \\ \cline{2-2}
\multicolumn{1}{|c|}{}                                       & \multicolumn{1}{c|}{$SVM$} & \multicolumn{1}{c}{\cellcolor[HTML]{FFC7CE}{\color[HTML]{9C0006} 0.1}} & \multicolumn{1}{c}{\cellcolor[HTML]{C6EFCE}{\color[HTML]{006100}  0}}   & \multicolumn{1}{c}{\cellcolor[HTML]{FFC7CE}{\color[HTML]{9C0006} 0.1}} & \multicolumn{1}{c}{\cellcolor[HTML]{FFEB9C}{\color[HTML]{9C5700} 0.1}} & \multicolumn{1}{c}{\cellcolor[HTML]{C6EFCE}{\color[HTML]{006100}  0}}   & \multicolumn{1}{c}{\cellcolor[HTML]{FFC7CE}{\color[HTML]{9C0006} 0.1}} & \multicolumn{1}{c}{\cellcolor[HTML]{C6EFCE}{\color[HTML]{006100}  0}}   & \multicolumn{1}{c}{\cellcolor[HTML]{C6EFCE}{\color[HTML]{006100}  0}}   & \multicolumn{1}{c}{\cellcolor[HTML]{FFC7CE}{\color[HTML]{9C0006} 0.3}} & \multicolumn{1}{c}{\cellcolor[HTML]{FFEB9C}{\color[HTML]{9C5700} 0.2}} & \multicolumn{1}{c}{\cellcolor[HTML]{FFEB9C}{\color[HTML]{9C5700} 0.4}} & \multicolumn{1}{c}{\cellcolor[HTML]{C6EFCE}{\color[HTML]{006100}  0}}   & \multicolumn{1}{c}{\cellcolor[HTML]{FFC7CE}{\color[HTML]{9C0006} 0.08}} & \multicolumn{1}{c}{\cellcolor[HTML]{FFC7CE}{\color[HTML]{9C0006} 0.08}} & \multicolumn{1}{c|}{\cellcolor[HTML]{FFC7CE}{\color[HTML]{9C0006} 0.08}} &  &  &  \\ \cline{2-2}
\multicolumn{1}{|c|}{\multirow{-3}{*}{PATRON (Average)}}      & \multicolumn{1}{c|}{$STVM_{sr}$}       & \multicolumn{1}{c}{\cellcolor[HTML]{FFC7CE}{\color[HTML]{9C0006}0.2}}                                                & \multicolumn{1}{c}{\cellcolor[HTML]{C6EFCE}{\color[HTML]{006100} 0}}                                                  & \multicolumn{1}{c}{\cellcolor[HTML]{FFC7CE}{\color[HTML]{9C0006}0.2}}                                                & \multicolumn{1}{c}{\cellcolor[HTML]{FFEB9C}{\color[HTML]{9C5700} 0}}   & \multicolumn{1}{c}{\cellcolor[HTML]{C6EFCE}{\color[HTML]{006100} 0} }                                                 & \multicolumn{1}{c}{\cellcolor[HTML]{FFC7CE}{\color[HTML]{9C0006}0.1}}                                               & \multicolumn{1}{c}{\cellcolor[HTML]{C6EFCE}{\color[HTML]{006100} 0}}                                                  & \multicolumn{1}{c}{\cellcolor[HTML]{C6EFCE}{\color[HTML]{006100} 0} }                                                 & \multicolumn{1}{c}{\cellcolor[HTML]{FFC7CE}{\color[HTML]{9C0006}0.2}}                                                & \multicolumn{1}{c}{\cellcolor[HTML]{FFEB9C}{\color[HTML]{9C5700} 0}}   & \multicolumn{1}{c}{\cellcolor[HTML]{FFEB9C}{\color[HTML]{9C5700} 0}}   & \multicolumn{1}{c}{\cellcolor[HTML]{C6EFCE}{\color[HTML]{006100} 0} }                                                 & \multicolumn{1}{c}{\cellcolor[HTML]{FFC7CE}{\color[HTML]{9C0006}0.1}}                                                 & \multicolumn{1}{c}{\cellcolor[HTML]{FFC7CE}{\color[HTML]{9C0006}0.1}}                                                 & \multicolumn{1}{c|}{\cellcolor[HTML]{FFC7CE}{\color[HTML]{9C0006}0.2}}                                                 &  &  &  \\ \cline{1-2}
\multicolumn{1}{|c|}{}                                       & \multicolumn{1}{c|}{$VM$}           & \multicolumn{1}{c}{\cellcolor[HTML]{C6EFCE}{\color[HTML]{006100} 0}}   & \multicolumn{1}{c}{\cellcolor[HTML]{C6EFCE}{\color[HTML]{006100} 0}}   & \multicolumn{1}{c}{\cellcolor[HTML]{C6EFCE}{\color[HTML]{006100} 0}}   & \multicolumn{1}{c}{\cellcolor[HTML]{C6EFCE}{\color[HTML]{006100} 0}}   & \multicolumn{1}{c}{\cellcolor[HTML]{C6EFCE}{\color[HTML]{006100} 0}}   & \multicolumn{1}{c}{\cellcolor[HTML]{C6EFCE}{\color[HTML]{006100} 0}}   & \multicolumn{1}{c}{\cellcolor[HTML]{C6EFCE}{\color[HTML]{006100} 0}}   & \multicolumn{1}{c}{\cellcolor[HTML]{C6EFCE}{\color[HTML]{006100} 0}}   & \multicolumn{1}{c}{\cellcolor[HTML]{C6EFCE}{\color[HTML]{006100} 0}}   & \multicolumn{1}{c}{\cellcolor[HTML]{C6EFCE}{\color[HTML]{006100} 0}}   & \multicolumn{1}{c}{\cellcolor[HTML]{C6EFCE}{\color[HTML]{006100} 0}}   & \multicolumn{1}{c}{\cellcolor[HTML]{C6EFCE}{\color[HTML]{006100} 0}}   & \multicolumn{1}{c}{\cellcolor[HTML]{C6EFCE}{\color[HTML]{006100} 0}}    & \multicolumn{1}{c}{\cellcolor[HTML]{C6EFCE}{\color[HTML]{006100} 0}}    & \multicolumn{1}{c|}{\cellcolor[HTML]{C6EFCE}{\color[HTML]{006100} 0}}    &  &  &  \\ \cline{2-2}
\multicolumn{1}{|c|}{}                                       & \multicolumn{1}{c|}{$SVM$} & \multicolumn{1}{c}{\cellcolor[HTML]{C6EFCE}{\color[HTML]{006100} 0}}   & \multicolumn{1}{c}{\cellcolor[HTML]{C6EFCE}{\color[HTML]{006100} 0}}   & \multicolumn{1}{c}{\cellcolor[HTML]{C6EFCE}{\color[HTML]{006100} 0}}   & \multicolumn{1}{c}{\cellcolor[HTML]{C6EFCE}{\color[HTML]{006100} 0}}   & \multicolumn{1}{c}{\cellcolor[HTML]{C6EFCE}{\color[HTML]{006100} 0}}   & \multicolumn{1}{c}{\cellcolor[HTML]{C6EFCE}{\color[HTML]{006100} 0}}   & \multicolumn{1}{c}{\cellcolor[HTML]{C6EFCE}{\color[HTML]{006100} 0}}   & \multicolumn{1}{c}{\cellcolor[HTML]{C6EFCE}{\color[HTML]{006100} 0}}   & \multicolumn{1}{c}{\cellcolor[HTML]{C6EFCE}{\color[HTML]{006100} 0}}   & \multicolumn{1}{c}{\cellcolor[HTML]{C6EFCE}{\color[HTML]{006100} 0}}   & \multicolumn{1}{c}{\cellcolor[HTML]{C6EFCE}{\color[HTML]{006100} 0}}   & \multicolumn{1}{c}{\cellcolor[HTML]{C6EFCE}{\color[HTML]{006100} 0}}   & \multicolumn{1}{c}{\cellcolor[HTML]{C6EFCE}{\color[HTML]{006100} 0}}    & \multicolumn{1}{c}{\cellcolor[HTML]{C6EFCE}{\color[HTML]{006100} 0}}    & \multicolumn{1}{c|}{\cellcolor[HTML]{C6EFCE}{\color[HTML]{006100} 0}}    &  &  &  \\ \cline{2-2}
\multicolumn{1}{|c|}{\multirow{-3}{*}{SPARSE (Average)}}      & \multicolumn{1}{c|}{$STVM_{sr}$}       & \multicolumn{1}{c}{\cellcolor[HTML]{C6EFCE}{\color[HTML]{006100} 0}}   & \multicolumn{1}{c}{\cellcolor[HTML]{C6EFCE}{\color[HTML]{006100} 0}}   & \multicolumn{1}{c}{\cellcolor[HTML]{C6EFCE}{\color[HTML]{006100} 0}}   & \multicolumn{1}{c}{\cellcolor[HTML]{C6EFCE}{\color[HTML]{006100} 0}}   & \multicolumn{1}{c}{\cellcolor[HTML]{C6EFCE}{\color[HTML]{006100} 0}}   & \multicolumn{1}{c}{\cellcolor[HTML]{C6EFCE}{\color[HTML]{006100} 0}}   & \multicolumn{1}{c}{\cellcolor[HTML]{C6EFCE}{\color[HTML]{006100} 0}}   & \multicolumn{1}{c}{\cellcolor[HTML]{C6EFCE}{\color[HTML]{006100} 0}}   & \multicolumn{1}{c}{\cellcolor[HTML]{C6EFCE}{\color[HTML]{006100} 0}}   & \multicolumn{1}{c}{\cellcolor[HTML]{C6EFCE}{\color[HTML]{006100} 0}}   & \multicolumn{1}{c}{\cellcolor[HTML]{C6EFCE}{\color[HTML]{006100} 0}}   & \multicolumn{1}{c}{\cellcolor[HTML]{C6EFCE}{\color[HTML]{006100} 0}}   & \multicolumn{1}{c}{\cellcolor[HTML]{C6EFCE}{\color[HTML]{006100} 0}}    & \multicolumn{1}{c}{\cellcolor[HTML]{C6EFCE}{\color[HTML]{006100} 0}}    & \multicolumn{1}{c|}{\cellcolor[HTML]{C6EFCE}{\color[HTML]{006100} 0}}    &  &  &  \\ \cline{1-2}
\multicolumn{1}{|c|}{}                                       & \multicolumn{1}{c|}{$VM$}           & \multicolumn{1}{c}{\cellcolor[HTML]{C6EFCE}{\color[HTML]{006100} 0}}   & \multicolumn{1}{c}{\cellcolor[HTML]{C6EFCE}{\color[HTML]{006100} 0}}   & \multicolumn{1}{c}{\cellcolor[HTML]{C6EFCE}{\color[HTML]{006100} 0}}   & \multicolumn{1}{c}{\cellcolor[HTML]{C6EFCE}{\color[HTML]{006100} 0}}   & \multicolumn{1}{c}{\cellcolor[HTML]{C6EFCE}{\color[HTML]{006100} 0}}   & \multicolumn{1}{c}{\cellcolor[HTML]{C6EFCE}{\color[HTML]{006100} 0}}   & \multicolumn{1}{c}{\cellcolor[HTML]{C6EFCE}{\color[HTML]{006100} 0}}   & \multicolumn{1}{c}{\cellcolor[HTML]{C6EFCE}{\color[HTML]{006100} 0}}   & \multicolumn{1}{c}{\cellcolor[HTML]{C6EFCE}{\color[HTML]{006100} 0}}   & \multicolumn{1}{c}{\cellcolor[HTML]{C6EFCE}{\color[HTML]{006100} 0}}   & \multicolumn{1}{c}{\cellcolor[HTML]{C6EFCE}{\color[HTML]{006100} 0}}   & \multicolumn{1}{c}{\cellcolor[HTML]{C6EFCE}{\color[HTML]{006100} 0}}   & \multicolumn{1}{c}{\cellcolor[HTML]{C6EFCE}{\color[HTML]{006100} 0}}    & \multicolumn{1}{c}{\cellcolor[HTML]{C6EFCE}{\color[HTML]{006100} 0}}    & \multicolumn{1}{c|}{\cellcolor[HTML]{C6EFCE}{\color[HTML]{006100} 0}}    &  &  &  \\ \cline{2-2}
\multicolumn{1}{|c|}{}                                       & \multicolumn{1}{c|}{$SVM$} & \multicolumn{1}{c}{\cellcolor[HTML]{C6EFCE}{\color[HTML]{006100}  0}}   & \multicolumn{1}{c}{\cellcolor[HTML]{C6EFCE}{\color[HTML]{006100}  0}}   & \multicolumn{1}{c}{\cellcolor[HTML]{FFC7CE}{\color[HTML]{9C0006} 0.1}} & \multicolumn{1}{c}{\cellcolor[HTML]{FFC7CE}{\color[HTML]{9C0006} 0.1}} & \multicolumn{1}{c}{\cellcolor[HTML]{C6EFCE}{\color[HTML]{006100}  0}}   & \multicolumn{1}{c}{\cellcolor[HTML]{C6EFCE}{\color[HTML]{006100}  0}}   & \multicolumn{1}{c}{\cellcolor[HTML]{C6EFCE}{\color[HTML]{006100}  0}}   & \multicolumn{1}{c}{\cellcolor[HTML]{FFEB9C}{\color[HTML]{9C5700} 0.4}} & \multicolumn{1}{c}{\cellcolor[HTML]{FFC7CE}{\color[HTML]{9C0006} 0.3}} & \multicolumn{1}{c}{\cellcolor[HTML]{FFEB9C}{\color[HTML]{9C5700} 0.2}} & \multicolumn{1}{c}{\cellcolor[HTML]{FFEB9C}{\color[HTML]{9C5700} 0.4}} & \multicolumn{1}{c}{\cellcolor[HTML]{C6EFCE}{\color[HTML]{006100}  0}}   & \multicolumn{1}{c}{\cellcolor[HTML]{FFEB9C}{\color[HTML]{9C5700} 0.08}} & \multicolumn{1}{c}{\cellcolor[HTML]{FFC7CE}{\color[HTML]{9C0006} 0.17}} & \multicolumn{1}{c|}{\cellcolor[HTML]{C6EFCE}{\color[HTML]{006100}  0}}    &  &  &  \\ \cline{2-2}
\multicolumn{1}{|c|}{\multirow{-3}{*}{Codetables (Average)}}  & \multicolumn{1}{c|}{$STVM_{sr}$}       & \multicolumn{1}{c}{\cellcolor[HTML]{C6EFCE}{\color[HTML]{006100} 0} }                                                 & \multicolumn{1}{c}{\cellcolor[HTML]{C6EFCE}{\color[HTML]{006100} 0}}                                                  & \multicolumn{1}{c}{\cellcolor[HTML]{FFC7CE}{\color[HTML]{9C0006}0.1}}                                                & \multicolumn{1}{c}{\cellcolor[HTML]{FFC7CE}{\color[HTML]{9C0006}0.1}}                                                & \multicolumn{1}{c}{\cellcolor[HTML]{C6EFCE}{\color[HTML]{006100} 0}}                                                  & \multicolumn{1}{c}{\cellcolor[HTML]{C6EFCE}{\color[HTML]{006100} 0}}                                                  & \multicolumn{1}{c}{\cellcolor[HTML]{C6EFCE}{\color[HTML]{006100} 0} }                                                 & \multicolumn{1}{c}{\cellcolor[HTML]{FFEB9C}{\color[HTML]{9C5700} 0}}   & \multicolumn{1}{c}{\cellcolor[HTML]{FFC7CE}{\color[HTML]{9C0006}0.2}}                                                & \multicolumn{1}{c}{\cellcolor[HTML]{FFEB9C}{\color[HTML]{9C5700} 0}}   & \multicolumn{1}{c}{\cellcolor[HTML]{FFEB9C}{\color[HTML]{9C5700} 0}}   & \multicolumn{1}{c}{\cellcolor[HTML]{C6EFCE}{\color[HTML]{006100} 0}}                                                  & \multicolumn{1}{c}{\cellcolor[HTML]{FFEB9C}{\color[HTML]{9C5700} 0}}    & \multicolumn{1}{c}{\cellcolor[HTML]{FFC7CE}{\color[HTML]{9C0006}0.23}}                                                & \multicolumn{1}{c|}{\cellcolor[HTML]{C6EFCE}{\color[HTML]{006100} 0}}                                                   &  &  &  \\ \cline{1-17}   
\end{tabular}%
}
\caption{Vulnerability metrics ($VM$, $SVM$, $STVM_{bf}, STVM_{sr}$) analysis.
Red and green rows denote vulnerable and non-vulnerable FSMs, respectively. For PATRON, SPARSE and Codetables, as 5 values are taken for each $x$ the maximum values of the corresponding vulnerability metrics are shown. The two vertically adjacent yellow cells highlight the occurrences of false positives by $SVM$ metric.}
\label{tableresults2}
\end{table*}

\vspace{0.5ex}
\noindent \textbf{FSM Security Resilience Comparison:} Note that, compared to SPARSE, TRANSPOSE provides more realistic vulnerability estimation owing to the whole design implementation in the layout (all $\mathbb{FF}$ corresponding to the whole design); SPARSE only considers the FSM encoded module.  
For security analysis, $VM$, $SVM$, $STVM_{bf}$, and $STVM_{sr}$ are explored with increasing $x$ as shown in Table \ref{tableresults2}. As the average of 5 values are taken for each $x$ for the state-based approaches, the maximum values of $SVM$ and $STVM_{sr}$ are noted for these approaches as security risk is to be assessed according to the worst-case-scenario.

Except for TRANSPOSE, all state-based methodologies inherently achieve an encoding where $VM=0$. In contrast, for all TRANSPOSE variations, $VM(x)>0$, indicating vulnerability of specific transitions to LFI despite $STVM_{bf/sr}=0$. For instance, TRANSPOSE (SHA-256, $x=[1,2,3]$) produces a spatially secure encoding for all $AT$ in the FSM, although $VM>0$. Essentially, $VM$ serves as a conservative metric because achieving a secure sense  (i.e., a value of 0), may require considering some non-critical state transitions as critical. Note that, although $VM>0$ in TRANSPOSE means $\{s_i \in \mathbb{SS} \:,  HD(s_i,s_j) \leq x,  \: s_j \in \mathbb{NS}\}$, i.e., states incorporated in ($\mathbb{T}-\mathbb{AT}$) may access the $\mathbb{SS}$, this approach removes overly constrained conditions to provide more efficient $n$ and enables reduction of switching activity to optimize power in the generated encoding. As expected,
$SVM$ and $STVM_{bf}$ behave in a similar way in terms of security. The reason $SVM=0$ for all TRANSPOSE approaches (set-reset models) is because 
of the manner appropriate placement between the $\mathbb{SFF}$ sensitive regions is ensured -- ICC2 places a nearby cell to not waste space which likely has dimension in multiples of $D$. So, had it been the exact inter-distance, vulnerability would probably manifest in terms of $STVM_{sr}$.

It is confirmed from analyzing TRANSPOSE (Bit-Flip), and SPARSE, where $VM=0$ signifies that not only all the $AT$ are secure spatially, but all the $T$ between the $SS$ and the $NS$ are \emph{conservatively} secure for SPARSE. 
Hence, the metric $STVM_{bf}$ can be concluded as a better predictor of vulnerability than $SVM$ as it considers protection of only the specific $\mathbb{AT}$ considering the layout unlike $SVM$. However, in terms of detecting vulnerability they are both equal, i.e., if $STVM_{bf}>0$ then $SVM>0$ and vice versa.

Among the state-based approaches, PATRON and Codetables cannot take set-reset model into account, as $STVM_{sr}>0$ is seen for some values. This means that in at least one of the 5 samples, the attacker is able to execute $\mathbb{AT}$ illegally according to the FF layout and laser position $(l_i)$; hence the encoding choices did not fulfill the security requirements as FF arrangement introduced vulnerability so that $f>1$. Despite $VM=0$ is achieved for these two approaches, security is still not ensured, which means the overhead addition in these approaches due to security measure is not beneficial. Note that, the occasional numerical differences in values in $SVM$ and $STVM_{sr}$ is due to the $|\mathbb{S}|$ and $|\mathbb{T}|$ difference in the highlighted vertically adjacent red cells.

In the highlighted vertically adjacent yellow cells, the occurrences of false positives as explained in Section~\ref{saotbabfm} is captured by the $SVM$ and $STVM_{sr}$ metrics. For $x=2$ in FSM Controller, we see such $\mathbb{FF}$ layout. As $\mathbb{S}=7$, $x=2$, and $\mathbb{SS}=4$, an $[n,d]$ linear code of $[6,3]$ may provide $\mathbb{SS}=\{000000, 000111, 110100, 110011\}$. If $\{FF_1, \cdots FF_6\}$ is used to represent the FF order, then post synthesis layout arrangement in ICC2 is seen to be 
$\mathbb{E}(x=2)=\{ \{\langle{FF_1}\rangle\}, \{\langle{FF_2}\rangle\}, \{\langle{FF_3}\rangle\}, \{\langle{FF_4}\rangle\}, \{\langle{FF_5}\rangle\}$, $\{\langle{FF_6}\rangle\}$, $\{\langle{FF_{5:6}}\}\}$ ($x=2$ means any two FFs combinations in curly brackets can be simultaneously considered). Hence, the $\mathbb{AT}=\{000000 \rightarrow 000111, 110100 \rightarrow 110011\}$ still remains secure despite $SVM>0$, because of $SVM$'s limited capability to only handle the bit-flip model. The occasional $STVM_{sr} = 0$ is derived from the chance selection of security-compliant encoding choices, i.e., the current state and next state of each authorized transition follow TRANSPOSE encoding and placement constraints. However, there is no guarantee that $STVM_{sr}$ will always be 0 in these approaches. Except for TRANSPOSE and SPARSE, none of the approaches can reliably generate encoding with $STVM_{sr}=0$. Note that, as SPARSE placement constraints are also overly conservative, i.e., the $\mathbb{SFF}$ are placed $>D$ distance apart instead of adjusting placement between only the $\mathbb{FF}$ sensitive regions it is expected that $STVM_{sr}=0$ and no vulnerability is found. Hence, for SPARSE even though the encoding may not be secured against set-reset model, the conservative placement constraints successfully  provides security, but at a higher cost than any of the TRANSPOSE approaches.

In summary, set only, reset only, and set and reset oriented approaches provide appropriate security with least overhead. Among them, the difference in overhead corresponds to the difference in the security transitions. The model unaware approaches (PATRON and Codetables) are incapable of accommodating the precise set-reset model; they only support the bit flip model, despite ensuring a minimum Hamming Distance of ($x+1$) between the codewords. \emph{The fact that $STVM_{bf}$ is more precise than $SVM$ and  $SVM \neq STVM_{sr}$ illustrates the need for TRANSPOSE which has the flexibility of protecting only the specific $AT$ considering the $\mathbb{FF}$ layout and both data-dependent and data-independent models in estimating the FSM vulnerability to LFI.}

\section{Conclusion}
In this paper, we introduced a spatially-aware transition-based encoding scheme resilient to LFI.
This scheme integrates FF placement and sensitive regions under bit set, reset, set and reset, and bit-flip models to safeguard any number and type of transitions in an FSM as specified by the designer.
Particularly, if the FF placement along with precise sensitive regions are unaccounted for in the threat model, critical errors result
for the contemporary countermeasures. 
In contrast, TRANSPOSE employs an automated LP approach that offers greater flexibility by co-optimizing FSM encoding, FF placement, taking into account precise FF-sensitive regions aligned with the technology node, diverse design specifications, and attacker capabilities. This holistic approach results in a single, power-optimized encoding.
The proposed spatial transitional vulnerability metrics demonstrated superior precision compared to other state exploration methods, particularly in fault detection accuracy across both data-dependent and independent models. TRANSPOSE consistently outperformed alternative FSM encoding schemes in terms of security, PDP, and area, often excelling in all three metrics. Future work aims to extend these concepts to Field-Programmable Gate Arrays (FPGAs).

\section*{Acknowledgments}
This research was funded by Intel and partially supported by the NSF under grant number 2117349.


\vspace{-25pt}
\begin{IEEEbiography}[{\includegraphics[width=1in,height=1.25in,clip,keepaspectratio]{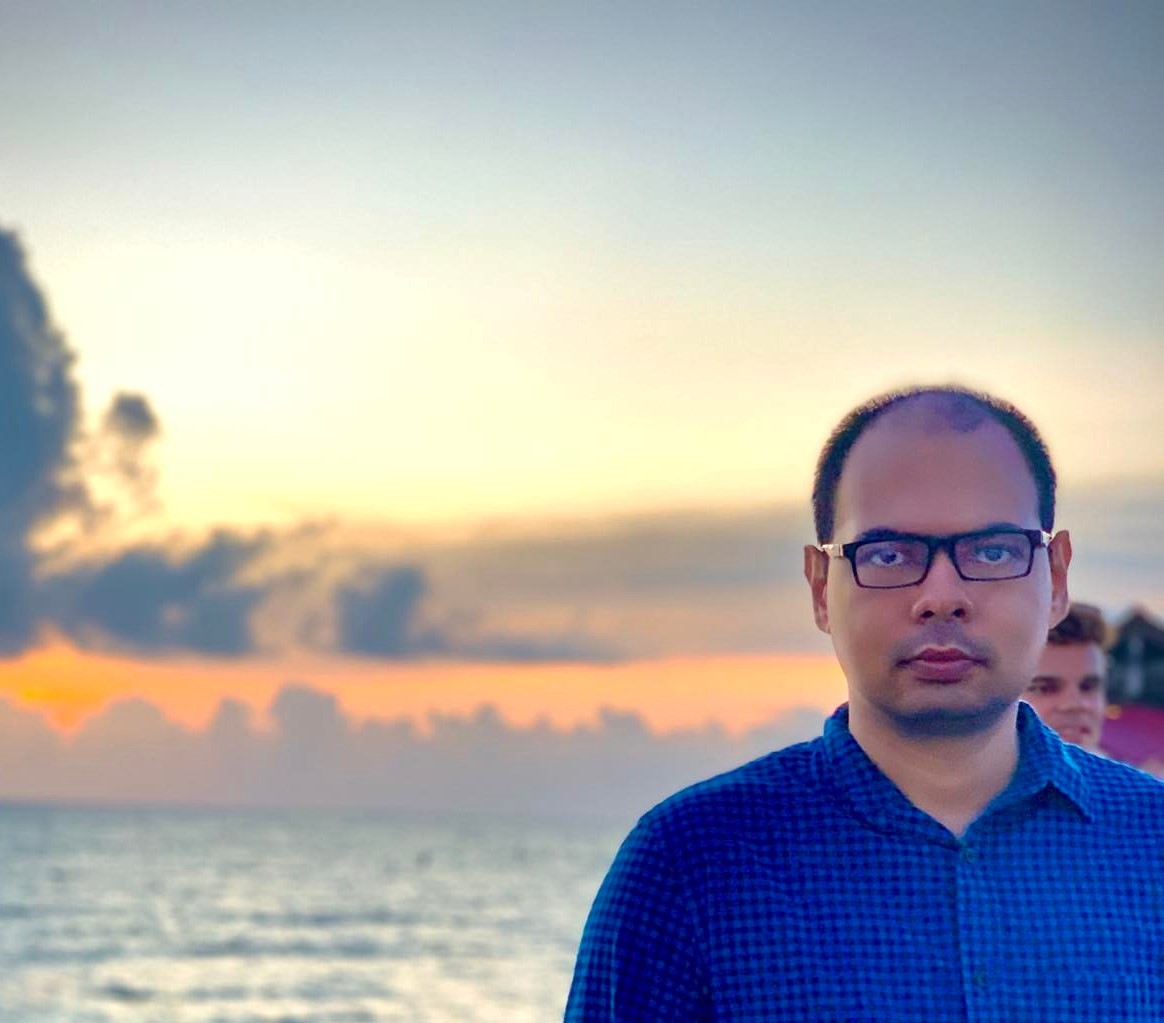}}]{MUHTADI CHOUDHURY}
received his doctorate in Electrical
and Computer Engineering Department, University
of Florida, Gainesville, FL, USA. His research is mainly 
focused on developing tools for vulnerability analysis and mitigation and developing and analyzing
security-aware designs against fault injection attacks. He received an M.S. degree from the University of Toledo, USA.
 \end{IEEEbiography}
\vspace{-25pt}
\begin{IEEEbiography}[{\includegraphics[width=1in,height=1.25in,clip,keepaspectratio]{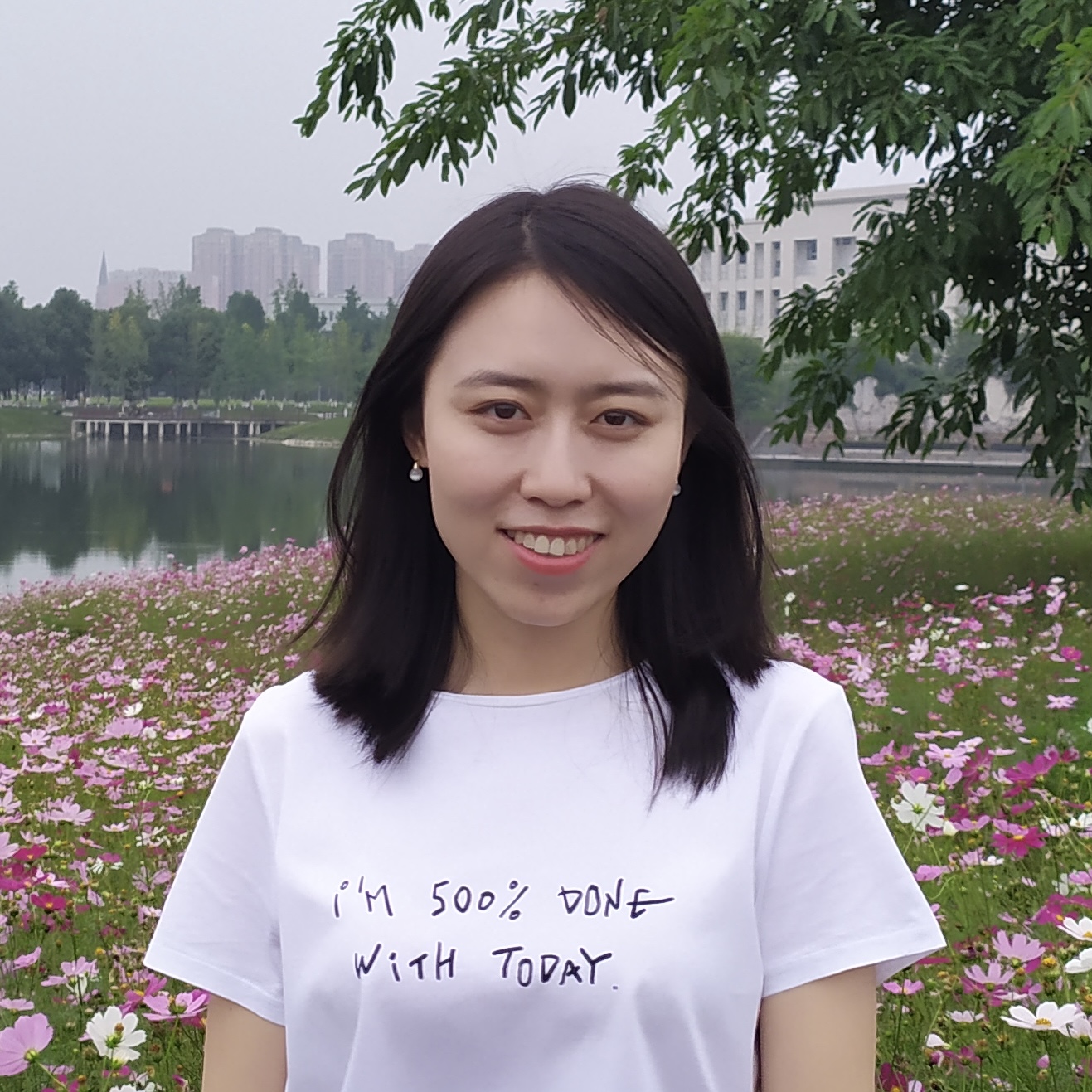}}]{MINYAN GAO}
received the B.S. degree from Northwest University, Xi’an, China, in 2016, and the M.E.
degree in Electrical Engineering from University of Virginia, Charlottesville, USA. She is currently working toward the Ph.D. degree in Electrical Engineering at University of Florida, Gainesville, FL, USA. Her current research interests include hardware security and trust, VLSI CAD, VLSI physical design.
 \end{IEEEbiography}
\vspace{-25pt}
\begin{IEEEbiography}[{\includegraphics[width=1in,height=1.25in,clip,keepaspectratio]{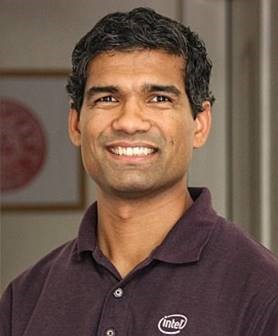}}]{Avinash Varna}
received the B.Tech. degree in electrical engineering from IIT Madras, Chennai, India, in 2005, and the Ph.D. degree in electrical engineering from the University of Maryland, College Park, MD, USA, in 2011. He is currently a Principal Engineer with Intel Corporation, Chandler, AZ, USA. His research interests include the security of embedded systems, applied cryptography, information forensics, and multimedia security. Dr. Varna served on the organizing committee of the IEEE International Workshop on Information Forensics and Security in 2014 and the IEEE Technical Committee on Information Forensics and Security from 2015 to 2017.
\end{IEEEbiography}
\vspace{-25pt}
\begin{IEEEbiography}[{\includegraphics[width=1in,height=1.25in,clip,keepaspectratio]{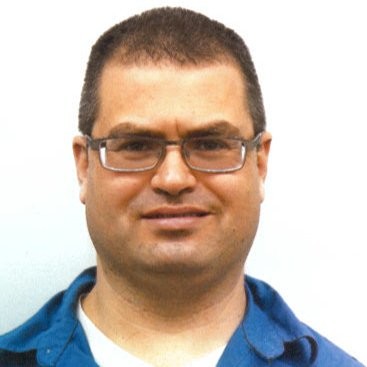}}]{Elad Peer}
is a security researcher at Intel corporation since 2016. Prior to that he worked with various companies including Cisco systems, NDS, Freescale semiconductors. His interest fields span hw security and reverse engineering, physical security, fault injection and side channels. He holds a PhD in nanotechnology from the Technion - Israel Institute of Technology (2012), an MSc in BioMedical engineering from Tel Aviv University, Israel (2005), and BSc in electrical and electronics engineering from the Technion-IIT (1995). Dr. Peer holds 4 patents and multiple publications in diverse fields.
\end{IEEEbiography}

\vspace{-25pt}

\begin{IEEEbiography}[{\includegraphics[width=1in,height=1.25in,clip,keepaspectratio]{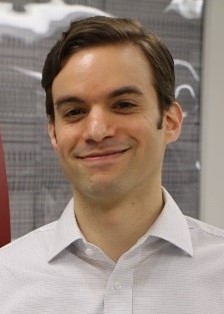}}]{DOMENIC FORTE}
received the B.S. degree
from the Manhattan College, Riverdale, NY, USA, and the M.S. and Ph.D. degrees from
the University of Maryland at College Park, College Park, MD, USA, respectively, all in electrical engineering. He is currently
an Associate Professor with the Electrical and
Computer Engineering Department, University of
Florida, Gainesville, FL, USA. His research interests
include the domain of hardware security, including
the investigation of hardware security primitives,
hardware Trojan detection and prevention, electronics supply chain security,
and anti-reverse engineering. 
\end{IEEEbiography}

\vfill

\end{document}